\documentclass[reprint,superscriptaddress,aps,twocolumn,showpacs]{revtex4-2}
\usepackage{amssymb}
\usepackage{amsmath}
\usepackage{graphicx}
\usepackage{multirow}
\RequirePackage{filecontents}
\usepackage{subcaption}
\allowdisplaybreaks
\usepackage{physics}
\usepackage{braket} 
\usepackage{hyperref}
\usepackage{xcolor}
\usepackage{slashed}
\usepackage{simpler-wick}
\usepackage{slashed}
\usepackage{bm}
\usepackage{makecell}
\usepackage{xcolor}
\usepackage[utf8]{inputenc}

\begin{document}

\preprint{APS/123-QED}

\title{Helicity amplitudes of $N(1520)$ and $N(1535)$ including pentaquark components}

\author{K. Xu}
\email{gxukai1123@gmail.com}
\address{School of Physics and Center of Excellence in High Energy Physics and Astrophysics, Suranaree University of Technology, Nakhon Ratchasima 30000, Thailand}
\author{A. Kaewsnod}
\author{Z. Zhao}
\author{A. Limphirat}
\email{ayut@g.sut.ac.th}
\author{Y. Yan}
\email{yupeng@g.sut.ac.th}
\address{School of Physics and Center of Excellence in High Energy Physics and Astrophysics, Suranaree University of Technology, Nakhon Ratchasima 30000, Thailand}
\date{\today}

\begin{abstract}
We investigate the helicity amplitudes of the $N(1520)$ and $N(1535)$ resonances in the electromagnetic transition process $\gamma^*p\to N^*$, where the $N(1520)$ and $N(1535)$ are assumed to include both the $L=1$ three-quark and $L=0$ spatial symmetric $q^4\bar q$ pentaquark components. The helicity transition amplitudes $A_{1/2}$, $A_{3/2}$ (for spin 3/2 states) and $S_{1/2}$ are computed within the constituent quark model. The inclusion of the $q^4 \bar q$ pentaquark components via the $\gamma^*\to q\bar q$ diagram significantly improves the theoretical description of the helicity amplitudes for both $N(1520)$ and $N(1535)$, yielding a closer agreement with experimental data compared to the pure three-quark picture. The work reveals that the $N(1520)$ and $N(1535)$ resonances may contain a considerable pentaquark components. 

\end{abstract}


\maketitle

\section{\label{sec1}Introduction}
In the past two decades, new data on the electroproduction of low-lying nucleon resonances which have been accumulated at JLab \cite{PhysRevLett.86.1702,PhysRevC.68.065204,PhysRevC.80.015205,PhysRevC.78.045209,PhysRevC.71.015201,Aznauryan:2009mx,PhysRevC.76.015204,PhysRevC.86.035203,PhysRevC.93.025206} and by the Mainz group \cite{Drechsel:2007if,Tiator:2009mt,Tiator:2011pw} provides us important information to reveal their inner structures. As the most common representation of inner structure, helicity amplitudes in all the $\gamma^* N \to N^*$ transitions are parametrized in terms of the three different photon polarizations, including two transverse amplitudes $A_{1/2}$ and $A_{3/2}$(for spin 3/2 states) and one longitudinal amplitude $S_{1/2}$. One may see some good reviews about helicity amplitudes in Refs. \cite{Drechsel2007,AZNAURYAN20121,RAMALHO2024104097}. 

Numerous theoretical approaches have been developed to describe the helicity amplitudes of $N(1520)$ and $N(1535)$, such as a variety of quark models: single quark transition model \cite{PhysRevC.67.035204}, light-front quark models \cite{Capstick:1994ne,PACE200033,PhysRevD.100.094013,PhysRevC.85.055202,PhysRevC.95.065207}, hypercentral constituent quark model \cite{Giannini:2015zia,Giannini:2001kb,PhysRevC.86.065202}, and QCD sum rules \cite{PhysRevLett.103.072001,PhysRevD.92.014018}. In particular, the electromagnetic properties of $N(1520)$, $N(1535)$, and several other nucleon resonances have been studied in the soft-wall Ads/QCD approaches \cite{PhysRevD.101.034026,PhysRevD.102.094008}.

Beyond the pure three-quark picture, more complex internal structures involving hadronic degrees of freedom have been explored. These include meson cloud effects, baryon-meson hadronic molecular components, and pentaquark components which all indicate a $qqq(q\bar q)$ quark content, have been proposed for a better understanding of the $\gamma^*p\to N(1520)/N(1535)$ transitions \cite{Burkert:2004sk,PhysRevC.88.035209,PhysRevC.77.065207,PhysRevC.93.025206,PhysRevC.85.055202,PhysRevD.89.094016,PhysRevD.95.014003,PhysRevD.85.093014,JuliaDiaz:2007fa,An2009,ZOU2010199}. For the N(1520) resonance, several studies suggest that baryon-meson molecular-like configurations could play a dominant role in its structure \cite{PhysRevC.93.025206,PhysRevD.89.094016,PhysRevD.95.014003}. Dynamical coupled-channel models further provide a unified framework in which both the $N(1520)$ and $N(1535)$ emerge naturally as baryon-meson resonant states \cite{Burkert:2004sk,PhysRevC.88.035209}. Similarly, within the unitary chiral model, $N(1535)$ is interpreted as a dynamically generated state with a strong baryon-meson component \cite{PhysRevC.77.065207}. And additional $qqq(q\bar q)$ components are incorporated into the wave functions of $N(1535)$ to improve the description of its helicity amplitudes, using distinct spatial scale parameters in Refs. \cite{An2009,ZOU2010199}.


In this work, we investigate the helicity amplitudes of the $N(1520)$ and $N(1535)$ resonances in a mixing structure of three-quark and pentaquark components, by extending a constituent quark model originally developed from the first principle to compute the helicity transition amplitudes of three-quark nucleon resonances within the quantum field theory framework \cite{PhysRevD.105.016008,MosEpja}. In Ref. \cite{MosEpja} we have pushed the three-quark contribution to its the upper limit, but it is found that only the $A_{1/2}$ of N(1520) and N(1535) can be fairly described. In the previous work \cite{Kai2020PRD} the $N(1535)1/2^-$ and $N(1520)3/2^-$ masses were described by including the ground-state pentaquark components into the three-quark configurations. This may suggest that both the $N(1520)$ and $N(1535)$ include a considerable pentaquark components. 

The paper is organized as follows. All possible quark configurations of ground state non-strange $q^4 \bar q$ pentaquarks are worked out in the language of permutation groups, the mass spectrum are also estimated in Sec.~\ref{sec2}. In Sec.~\ref{sec3} we briefly introduce the formalism of helicity amplitudes $A_{1/2}$, $A_{3/2}$, and $S_{1/2}$ of the $N(1520)$ and $N(1535)$ resonances in the photoproduction transitions $\gamma^* p \to N^*$, where both the three-quark and pentaquark configurations are considered. The helicity amplitudes of the $N(1520)$ and $N(1535)$ are calculated in the conventional three-quark picture and the mixing picture of three-quark and pentaquark components in the range $0\leq Q^2 \leq4.5 \;{\rm GeV^2}$ and fitted to the experimental data in Sec.~\ref{sec4}. Finally a discussion of the results and conclusions are given in Sec. \ref{sec5}.

\section{\label{sec2}$q^4 \bar q$ pentaquark states}
\indent The construction of pentaquark wave functions has been described in a group theory approach in Refs.~\cite{Kai2014,PhysRevC.100.065207,Kai2020PRD}, and the masses of light-quark and charmonium-like pentaquark states are calculated in a non-relativistic quark model. In this section, for the completeness of the article and the convenience to audiences, we present the main points and results of Refs. ~\cite{Kai2014,PhysRevC.100.065207,Kai2020PRD}.

\indent The construction of $q^4 \bar q$ states follow the rules that a $q^4 \bar q$ state must be a color
singlet and the $q^4 \bar q$ wave function should be
antisymmetric under any permutation between identical quarks. The pentaquark should be a color singlet demands that the color part of the pentaquark wave function must be a $[222]_1$ singlet. Since the color part of the antiquark in pentaquark states is a $[11]_3$ anti-triplet, the color wave function of the four-quark configuration must be a $[211]_3$ triplet. The total wave function of the four quark configuration is antisymmetric implies that its spatial-spin-flavor part must be a $[31]$ state by conjugation. 


\indent The total wave function of the $q^4$ configuration may be written in
the general form,
\begin{eqnarray}\label{eq::1}
\Psi_{total}&=&\sum_{i,j=\rho,\lambda,\eta}a_{ij}\; \psi^c_{[211]_i}\psi^{osf}_{[31]_j},
\end{eqnarray}
with
\begin{eqnarray}\label{eq::2}
\psi^{osf}_{[31]_{\rho,\lambda,\eta}} &=& \sum_{i,j=S,A,\rho,\lambda,\eta}b_{ij}\psi^{o}_{[X]_{i}}\psi^{sf}_{[Y]_{j}}, \nonumber \\
\psi^{sf}_{[Y]} &=& \sum_{i,j=S,A,\rho,\lambda,\eta}c_{ij}\psi^{s}_{[x]_{i}}\psi^{f}_{[y]_{j}},
\end{eqnarray}
where $\psi^c$, $\psi^{osf}$, $\psi^{sf}$, $\psi^{s}$ and $\psi^{f}$ are respectively the color, spatial-spin-flavor, spin-flavor, spin and flavor parts of the $q^4$ cluster.
$S,\;A,\;\rho,\;\lambda,\;\eta$ stand for fully symmetric, fully antisymmetric, $\rho$-type, $\lambda$-type and $\eta$-type functions.
The coefficients in Eqs. (\ref{eq::1}) and (\ref{eq::2}) can be determined by acting the permutations $(12)$, $(23)$ and $(34)$ of the $S_4$ group on both sides of the general wave functions.
The fully antisymmetric wave function for the $q^4$ cluster is worked out,
\indent
\begin{eqnarray}\psi = \frac{1}{\sqrt{3}}
\left( \psi^{c}_{[211]_\lambda} \psi^{osf}_{[31]_\rho} -
\psi^{c}_{[211]_\rho} \psi^{osf}_{[31]_\lambda} +
\psi^{c}_{[211]_\eta} \psi^{osf}_{[31]_\eta} \right)
\end{eqnarray}


\begin{table}[t]
\caption{\label{tab:configuration31}$[31]_{FS}$ spin-flavor configurations}
\begin{ruledtabular}
		\begin{tabular}{lll}
				\multicolumn{3}{c}{$[31]_{FS}$}\\
				\hline
				\multicolumn{1}{l}{$[31]_{FS}[31]_{F}[22]_{S}$}&
				\multicolumn{1}{l}{$[31]_{FS}[31]_{F}[31]_{S}$}&
				\multicolumn{1}{l}{$[31]_{FS}[31]_{F}[4]_{S}$}\\
				\multicolumn{1}{l}{$[31]_{FS}[211]_{F}[22]_{S}$}&
				\multicolumn{1}{l}{$[31]_{FS}[211]_{F}[31]_{S}$}&
				\multicolumn{1}{l}{$[31]_{FS}[22]_{F}[31]_{S}$}\\
				\multicolumn{1}{l}{$[31]_{FS}[4]_{F}[31]_{S}$}\\
	\end{tabular}
\end{ruledtabular}
\end{table}

For $L=0$ spatial symmetric pentaquark states, both spatial-spin-flavor and spin-flavor wave functions are in the $[31]$ configuration, and all possible spin and flavor configurations are shown in Table \ref{tab:configuration31}. As an example, the explicit spin-flavor wave functions of $[31]_{FS}[22]_{F}[31]_{S}$ for the $q^4$ cluster are derived as,
\begin{eqnarray}\label{d}
\psi^{\rm sf}_{[31]_\lambda} &=& -\frac{1}{2} \phi_{[22]_{\rho}}
\chi_{[31]_{\rho}} +\frac{1}{2} \phi_{[22]_{\lambda}}
\chi_{[31]_{\lambda}} +\frac{1}{\sqrt{2}} \phi_{[22]_{\lambda}}
\chi_{[31]_{\eta}}
\nonumber\\
\psi^{\rm sf}_{[31]_\rho} &=& -\frac{1}{2}
\phi_{[22]_{\rho}} \chi_{[31]_{\lambda}} -\frac{1}{2}
\phi_{[22]_{\lambda}} \chi_{[31]_{\rho}} +\frac{1}{\sqrt{2}}
\phi_{[22]_{\rho}} \chi_{[31]_{\eta}}
\nonumber\\
\psi^{\rm sf}_{[31]_\eta} &=&
 \frac{1}{\sqrt{2}} \phi_{[22]_{\rho}} \chi_{[31]_{\rho}}
+\frac{1}{\sqrt{2}} \phi_{[22]_{\lambda}} \chi_{[31]_{\lambda}}
\end{eqnarray}

The spatial wave functions of the $q^4\bar q$ pentaquark states may be expanded  in the complete basis constructed from the harmonic oscillator wave functions, which take the general form, 
\begin{table*}[!htb]
\caption{\label{ps1}Normalized $q^4\bar q$ symmetric-type spatial wave functions.}
\begin{ruledtabular}
\begin{tabular}{@{}ll@{}}
$\Psi_1^{[5]_S}$ & $\psi_{000}(\bm\rho)\psi_{000}(\bm\lambda)\psi_{000}(\bm\eta)\psi _{000}(\bm\xi )$ \\ [2pt]
$\Psi_2^{[5]_S}$ & $\frac{1}{\sqrt{3}}[\psi_{100}(\bm\rho)\psi_{000}(\bm\lambda)\psi_{000}(\bm\eta)\psi _{000}(\bm\xi )
                                +\psi_{000}(\bm\rho)\psi_{100}(\bm\lambda)\psi_{000}(\bm\eta)\psi _{000}(\bm\xi )+\psi_{000}(\bm\rho)\psi_{000}(\bm\lambda)\psi_{100}(\bm\eta)\psi _{000}(\bm\xi)]$  \\ [2pt]
$\Psi_3^{[5]_S}$ & $\psi_{000}(\bm\rho)\psi_{000}(\bm\lambda)\psi_{000}(\bm\eta)\psi _{100}(\bm\xi )$\\[2pt]
$\Psi_4^{[5]_S}$ & $\sqrt{\frac{5}{33}}[\psi_{200}(\bm\rho)\psi_{000}(\bm\lambda)\psi_{000}(\bm\eta)
                                +\psi_{000}(\bm\rho)\psi_{200}(\bm\lambda)\psi_{000}(\bm\eta)+\psi_{000}(\bm\rho)	\psi_{000}(\bm\lambda)\psi_{200}(\bm\eta)$  \\ 
$$   &  $+\sqrt{\frac{6}{5}}\psi_{100}(\bm\rho)\psi_{100}(\bm\lambda)\psi_{000}(\bm\eta)+\sqrt{\frac{6}{5}}\psi_{100}(\bm\rho)\psi_{000}(\bm\lambda)\psi_{100}(\bm\eta)+\sqrt{\frac{6}{5}}\psi_{000}(\bm\rho)\psi_{100}(\bm\lambda)\psi_{100}(\bm\eta)] \psi _{000}(\bm\xi )$  \\[2pt]                           
$\Psi_5^{[5]_S}$ & $\frac{1}{\sqrt{3}}[\psi_{100}(\bm\rho)\psi_{000}(\bm\lambda)\psi_{000}(\bm\eta)\psi _{100}(\bm\xi )
                                +\psi_{000}(\bm\rho)\psi_{100}(\bm\lambda)\psi_{000}(\bm\eta)\psi _{100}(\bm\xi)+\psi_{000}(\bm\rho)\psi_{000}(\bm\lambda)\psi_{100}(\bm\eta)\psi _{100}(\bm\xi )]$  \\ [2pt]
$\Psi_6^{[5]_S}$ & $\psi_{000}(\bm\rho)\psi_{000}(\bm\lambda)\psi_{000}(\bm\eta)\psi _{200}(\bm\xi)$ \\ [2pt]
\end{tabular}
\end{ruledtabular}
\end{table*}

\begin{eqnarray}\label{eqn::spatial2}
\Psi_{NLM}^{q^4 \bar q[X]_y}&=& \sum_{\{n_i,l_i\}}
 A(n_{\rho},n_{\lambda},n_{\eta},n_{\xi},l_{\rho},l_{\lambda},l_{\eta}, l_\xi) \nonumber \\ 
&& \times \psi_{n_{\rho}l_{\rho}}(\vec\rho\,)\otimes\psi_{n_{\lambda}l_{\lambda}}(\vec\lambda\,) \otimes\psi_{n_{\eta}l_{\eta}}(\vec\eta\,)\otimes \psi_{n_\xi,l_\xi}(\vec\xi)
\nonumber \\
&=& \sum_{\{n_i,l_i,m_i\}} C_{n_{\rho}, l_{\rho},m_{\rho},n_{\lambda},l_{\lambda},m_{\lambda},n_{\eta},l_{\eta},m_{\eta},n_{\xi},l_{\xi},m_{\xi}} \nonumber \\
&& \times \psi_{n_{\rho}l_{\rho}m_{\rho}}(\vec\rho\,)\psi_{n_{\lambda}l_{\lambda}m_{\lambda}}(\vec\lambda\,)\psi_{n_{\eta}l_{\eta}m_{\eta}}(\vec\eta\,)\psi_{n_{\xi}l_{\xi}m_{\xi}}(\vec\xi) \nonumber \\
\end{eqnarray}
with $\psi_{n_{i}l_{i}m_{i}}$ being harmonic oscillator wave functions in the Jacobi coordinates $\rho,\lambda,\eta$, and $\xi$,
\begin{eqnarray}\label{eq::14}
	\vec{\rho} &=& \frac{1}{\sqrt{2}}(\vec{r_{1}}-\vec{r_{2}})\nonumber \\
	\vec{\lambda} &=& \frac{1}{\sqrt{6}}(\vec{r_{1}}+\vec{r_{2}}-2\vec{r_{3}})
	\nonumber \\
	\vec{\eta}&=&\frac{1}{\sqrt{12}}(\vec{r_{1}}+\vec{r_{2}}+\vec{r_{3}}-3\vec{r_{4}})
	\nonumber \\
	\vec{\xi}&=&\frac{1}{\sqrt{20}}(\vec{r_{1}}+\vec{r_{2}}+\vec{r_{3}}+\vec{r_{4}}-4\vec{r_{5}})
\end{eqnarray} 
where the antiquark is assigned the coordinate $\vec r_5$.
$N$, $L$, and $M$ are respectively the total principle quantum number, total angular momentum and magnetic quantum number of the pentaquark, with
\begin{eqnarray}\label{eqn::q4xi}
N= 2n_{\rho}+ l_{\rho}+2n_{\lambda}+l_{\lambda}+2n_{\eta}+ l_{\eta}+2n_{\xi}+l_{\xi}
\end{eqnarray}
$[X]_{y}$ in Eq. (\ref{eqn::spatial2}) stand for all possible permutation symmetries of the $q^4$ cluster, that is, $[X]_{y} = [4]_S,\;[31]_{\rho,\lambda,\eta},\;[211]_{\rho,\lambda,\eta},\;[22]_{\rho,\lambda}$. Coupling the antiquark to the $q^4$ subsystem preserves the permutation symmetry properties of the $q^4$ component within the $q^4\bar q$ pentaquark states, since $\psi_{n_\xi,l_\xi}(\vec\xi)$ is fully symmetric for any permutation between quarks.

In principle, one can construct the harmonic oscillator basis for pentaquark states to any order by applying the permutations symmetry of the $q^4 \bar q$ system to the general form in Eq. (\ref{eqn::spatial2}). The explicit normalized harmonic oscillator basis up to $N=4$ are listed in Table \ref{ps1}, where $l_\rho$, $l_\lambda$, $l_\eta$, and $L$ are limited to $0$ only.
\begin{table}[b]
\caption{\label{tab:d1}$q^4\bar q$ ground state pentaquark masses.}
\begin{ruledtabular}
\begin{tabular}{lccc}
\multicolumn{1}{l}{$J^P$}&
 \multicolumn{1}{c}{$q^4 \bar q$ configurations}& 
 \multicolumn{1}{c}{$(S^{q^4}, S^{\bar q}, S)$}& 
 \multicolumn{1}{c}{$M^{EV}(q^4\bar q)$ ({\rm MeV})} \\
\hline
$\frac{5}{2}^{-}$  & $\Psi^{sf}_{[31]_{F}[4]_{S}}(q^4\bar q)$ & (2,1/2,5/2) &  2269
\\
\hline
$\frac{3}{2}^{-}$  & $\Psi^{sf}_{[4]_{F}[31]_{S}}(q^4\bar q)$ & (1,1/2,3/2) &  2269
\\
$\phantom{-}$ & $\begin{pmatrix} \Psi^{sf}_{[31]_{F}[4]_{S}}(q^4 \bar q) \\ \Psi^{sf}_{[31]_{F}[31]_{S}}(q^4 \bar q) \end{pmatrix}$ & \makecell[c]{(2,1/2,3/2) \\ (1,1/2,3/2)} &  $\begin{pmatrix}1805  \\2269 \end{pmatrix}$
\\
$\phantom{-}$ & $\Psi^{sf}_{[22]_{F}[31]_{S}}(q^4\bar q)$ & (1,1/2,3/2)  & 2049
\\
\hline
$\frac{1}{2}^{-}$ & $\Psi^{sf}_{[4]_{F}[31]_{S}}(q^4\bar q)$ & (1,1/2,1/2)  & 2562
\\
$\phantom{-}$ & $\begin{pmatrix}\Psi^{sf}_{[31]_{F}[31]_{S}}(q^4 \bar q)  \\ \Psi^{sf}_{[31]_{F}[22]_{S}}(q^4 \bar q) \end{pmatrix}$  & \makecell[c]{(1,1/2,1/2)  \\(0,1/2,1/2)}   & $\begin{pmatrix}
1986  \\ 2162 \end{pmatrix}$ 
\\
$\phantom{-}$ & $\Psi^{sf}_{[22]_{F}[31]_{S}}(q^4\bar q)$ & (1,1/2,1/2)  & 1683
\\
\end{tabular}
\end{ruledtabular}
\end{table}

\indent Then, we calculate the $q^4\bar q$ pentaquark masses in the Hamiltonian \cite{PhysRevC.100.065207,Kai2020PRD}, 
\begin{flalign}\label{eqn::ham}
&H =H_0+ H_{hyp}^{OGE}, \nonumber \\
&H_{0} =\sum_{k=1}^{N} (m_k+\frac{p_k^2}{2m_{k}})+\sum_{i<j}^{N}(-\frac{3}{8}\lambda^{C}_{i}\cdot\lambda^{C}_{j})(A_{ij} r_{ij}-\frac{B_{ij}}{r_{ij}}),  \nonumber \\
&H_{hyp}^{OGE} = -C_{OGE}\sum_{i<j}\frac{\lambda^{C}_{i}\cdot\lambda^{C}_{j}}{m_{i}m_{j}}\,\vec\sigma_{i}\cdot\vec\sigma_{j},
\end{flalign}
where $A_{ij}$ and $B_{ij}$ are mass-dependent coupling parameters, taking the form
\begin{eqnarray}
A_{ij}= a \sqrt{\frac{m_{ij}}{m_u}},\;\;B_{ij}=b \sqrt{\frac{m_u}{m_{ij}}}.
\end{eqnarray}
with $m_{ij}$ being the reduced mass of ith and jth quarks, defined as $\;m_{ij}=\frac{2 m_i m_j}{m_i+m_j}$ which corresponds to the relative Jacobi coordinates of two-body system, $C_{OGE} = C_m\,m_u^2$, with $m_u$ being the constituent $u$ quark mass and $C_m$ a constant. $\lambda^C_{i}$ in the above equations are the generators of color SU(3) group. 

The three model coupling constants and four constituent quark masses are fitted to the low-lying baryon and ground heavy baryon spectra,
\begin{eqnarray}\label{eq:nmo}
&
m_u = m_d = 327 \ {\rm MeV}\,, \quad
m_s = 498 \ {\rm MeV}\,, \nonumber\\
&
m_c = 1642 \ {\rm MeV}\,, \quad
m_b = 4960 \ {\rm MeV}\,, \nonumber\\
&
C_m   =  18.3 \ {\rm MeV}, \quad
a     = 49500 \ {\rm MeV^2}, \quad
b     =  0.75 \nonumber\\
\end{eqnarray}

The mass spectra of the ground state $q^4\bar q$ pentaquarks are evaluated by solving the Schr\"odinger equation in the Hamiltonian in Eq. (\ref{eqn::ham}), including the mixtures of the configurations of the identical flavor parts.
The results are presented in Table \ref{tab:d1}. It is predicted in the calculation that the pentaquark state with the ${[31]_{FS}[22]_{F}[31]_{S}}$ configuration and the quantum numbers $I(J^P)=\frac1{2}(\frac1{2}^{-})$ has the lowest mass, around 1680 ${\rm MeV}$.


\section{\label{sec3}Formalism of helicity amplitudes in $q^3_{L=1}$ and $q^4\bar q_{L=0}$ mixing picture}

\begin{figure}[!t]
	\begin{center}
		\includegraphics[width=0.45\textwidth]{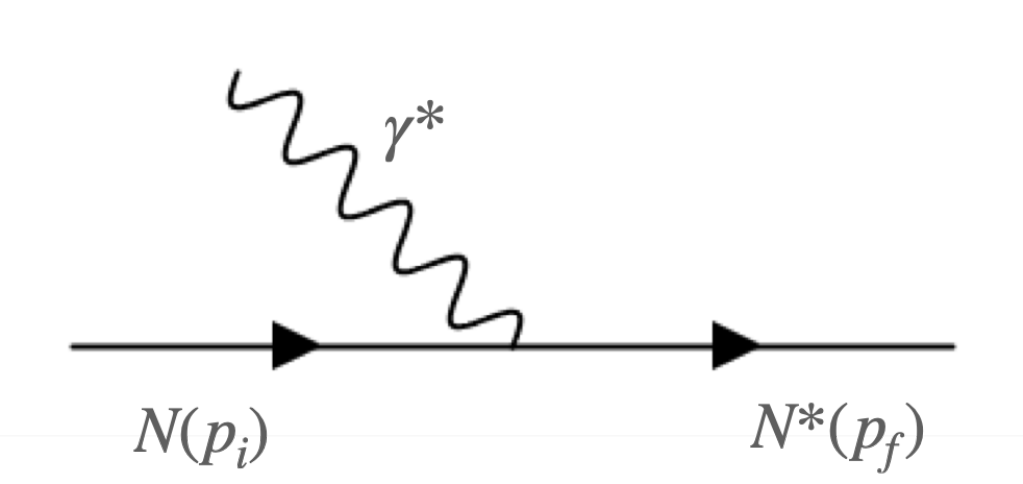}
		\caption{$\gamma^* p \to N^*$ process.}\label{photop}
	\end{center}
\end{figure}

We study the helicity amplitudes in the photo-production transitions $\gamma^* N\rightarrow N(1520)/N(1535)$, as shown in Fig. \ref{photop}, where both the $N(1520)3/2^-$ and $N(1535)1/2^-$ resonances ($N^*$) are considered in the mixing pictures of $L=1$ three-quark configuration and $L=0$ spatial symmetric $q^4\bar q$ pentaquark components,
\begin{eqnarray}\label{eq:ngam}
	|N^*, J\rangle &=& B_{3q}  |q^3, L=1, J \rangle +B_{5q} |q^4\bar q, L=0, J\rangle 
\end{eqnarray}
with $B_{3q}^2 + B_{5q}^2 = 1$ and $J$=1/2 for $N(1535)$ and $J$=3/2 for $N(1520)$.
The momentum and mass of the initial $N$ and final $N^*$ in Fig. \ref{photop} are denoted as ($p_i$, $M_N$) and ($p_f$, $M_{N^*}$), respectively. The helicity amplitudes are defined in the $N^*$ rest frame, where the square of the four-momentum transfer is expressed by $Q^2=-k^2=|\bm k|^2-\omega^2$ with the photon $k=(\omega, 0, 0, |\bm k|$). The momentum conservation reads, $p_f=p_i+k$. One has
\begin{alignat}{2}\label{eq:pphoton}
       &\omega&&= \frac{M_{N^*}^2-M_N^2-Q^2}{2M_{N^*}},\nonumber\\
	&|\bm k|&&= \left[Q^2+(\frac{M_{N^*}^2-M_N^2-Q^2}{2M_{N^*}})^2\right]^\frac{1}{2}.
\end{alignat}
 The general helicity amplitudes $A_{1/2}$, $A_{3/2}$ and $S_{1/2}$ for $N(1520)$ and $N(1535)$ consist of the contributions from both the three-quark and pentaquark components, according to Eq. (\ref{eq:ngam}).
 
The helicity amplitudes of $N(1520)$ and $N(1535)$ in the $L=1$ three-quark picture has been evaluated within the approach in Refs. \cite{PhysRevD.105.016008,MosEpja}, using the impulse approximation,
 
 \begin{alignat}{3}\label{eq:helicity3q}
&A^{(3q)}_{i}&=&\ \dfrac{1}{\sqrt{2K}}\Braket{N^*,J'_z=i|q_1'q_2'q_3'}\nonumber\\
&&\times& T^{+}(q_1q_2q_3\rightarrow q_1'q_2'q_3')\Braket{q_1q_2q_3|N,J_z=i-1},\nonumber\\
&S^{(3q)}_{1/2}&=&\ \dfrac{1}{\sqrt{2K}}\Braket{N^*,J'_z=\frac{1}{2}|q_1'q_2'q_3'}\nonumber\\
&&\times& T^{0}(q_1q_2q_3\rightarrow q_1'q_2'q_3')\Braket{q_1q_2q_3|N,J_z=\frac{1}{2}}
\end{alignat}

The wave functions of the proton, and the $N(1520)$ and $N(1535)$ resonances in the $q^3$ configuration are constructed in group theory approach \cite{Kai2014,PhysRevC.100.065207,Kai2020PRD}, and their spatial wave functions are expanded in the harmonic oscillator basis in the three-quark picture \cite{PhysRevD.105.016008,MosEpja}. 

\begin{figure}[!t]
\begin{center}
\includegraphics[width=0.45\textwidth]{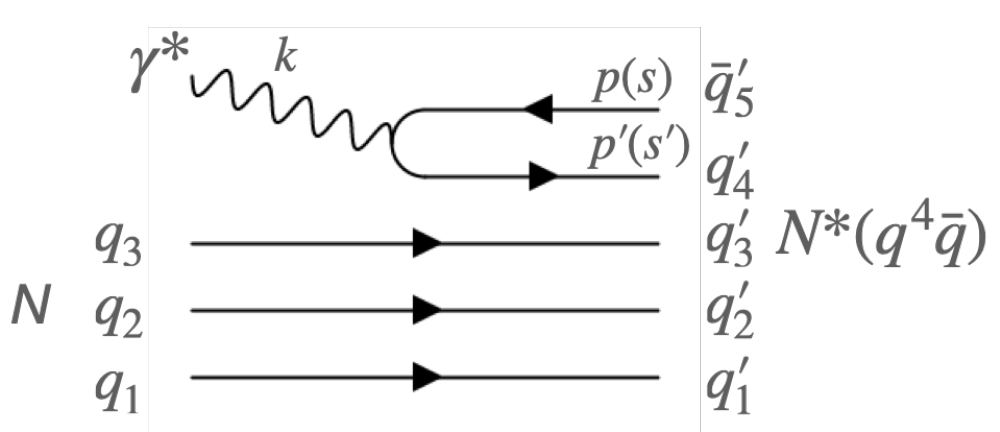}
\caption{Diagram of $\gamma^* p \to N^*(q^4\bar q)$ process via $\gamma^*\to q\bar q$ mechanism.}\label{pcdiagram}
\end{center}
\end{figure}

Then the helicity amplitudes of $N^*$ in the pentaquark picture with $L=0$ symmetric spatial wave functions are derived through $\gamma^*\to q\bar q$ mechanism shown in the Feynman diagram in Fig. \ref{pcdiagram}. This mechanism has been applied in Refs. \cite{An2009,ZOU2010199} to account for the pentaquark component and give a proper description of $A_{1/2}$ of $N(1535)$. The helicity amplitudes $A_{1/2}$, $A_{3/2}$ and $S_{1/2}$ in the pentaquark picture are,

\begin{eqnarray}\label{eq:helicity5q}
A^{(5q)}_{i}&=&\ \dfrac{1}{\sqrt{2K}}\Braket{N^*,J'_z=i|q_1'q_2'q_3'q'_4\bar {q'_5}} \nonumber \\
&\times& T^{+}(q_1q_2q_3 \gamma^* \rightarrow q_1'q_2'q_3'q'_4\bar {q'_5})\Braket{q_1q_2q_3 |N,J_z=i-1}, \nonumber\\
S^{(5q)}_{1/2}&=&\ \dfrac{1}{\sqrt{2K}}\Braket{N^*,J'_z=1/2|q_1'q_2'q_3'q'_4\bar {q'_5}} \nonumber\\
&\times& T^{0}(q_1q_2q_3 \gamma^* \rightarrow q_1'q_2'q_3'q'_4\bar {q'_5})\Braket{q_1q_2q_3 |N,J_z=1/2} \nonumber\\
\end{eqnarray}
with 
\begin{alignat}{2}\label{eq:pphotonon}
      	&K&&= \frac{M_{N^*}^2-M_N^2}{2M_{N^*}}.
\end{alignat}
where $i=\frac{1}{2}$ for $A_{1/2}$ and $i=\frac{3}{2}$ for $A_{3/2}$, and $K$ in Eq. (\ref{eq:pphoton}) is the real-photon momentum in the $N^*$ rest frame. $\Braket{q_1q_2q_3|N,J_z}$ in Eqs. (\ref{eq:helicity3q}) and (\ref{eq:helicity5q}) are the proton wave function, $\Braket{q_1'q_2'q_3'|N^*,J'_z}$ in Eq. (\ref{eq:helicity3q}) and $\Braket{q_1'q_2'q_3'q'_4\bar{q'_5}|N^*,J'_z}$ in Eq. (\ref{eq:helicity5q}) are the wave functions of $L=1$ three-quark and $L=0$ pentaquark states respectively. The transition amplitudes $T(q_1q_2q_3\rightarrow q_1'q_2'q_3'q'_4\bar {q'_5})$ through the $\gamma^*\to q\bar q$ process, as displayed in Fig. \ref{pcdiagram}, are calculated in the standard relativistic approach,
\begin{eqnarray}\label{transitionapp}
	&&T^{\lambda}(q_1q_2q_3 \gamma^* \rightarrow q_1'q_2'q_3'q'_4\bar {q'_5}) \nonumber\\
	&=& 4\,e_q\, \bar u_{s'} (p') \gamma^\mu \nu_{s}(p) \epsilon^\lambda_\mu(k)\Braket{q'_1q'_2q'_3|q_1q_2q_3} \nonumber\\
	&=& 4\, e_q\, T_{s's}^\lambda\Braket{q'_1q'_2q'_3|q_1q_2q_3}.
\end{eqnarray}
where $u_{s'}$ and $v_{s}$ are the Dirac spinors of the corresponding quark and antiquark, $e_q$ is the electric charge of the quark, and $\lambda$ is the helicity of the photon. A factor of 4 is multiplied, considering the permutation symmetry of the four constituent quarks. The photon polarization vectors $\epsilon_\mu^\lambda(k)$ take $\epsilon^0_\mu=(1,0,0,0)$ and $\epsilon^+_\mu=-\frac{1}{\sqrt{2}}(0,1,i,0)$. $\Braket{q'_1q'_2q'_3|q_1q_2q_3}$ leads to three $\delta$ functions, and thus the helicity amplitudes, $T^{\lambda}$ are dominated by the transition amplitudes $T_{s's}^\lambda$ in the quark-antiquark pair creation process $\gamma^*\to q\bar q$. The explicit matrix elements $T_{s's}^\lambda$ for different spins of the quark-antiquark pair and the helicity $\lambda=0, +1$, are presented in Eq. (\ref{matrixelement1}) in Appendix \ref{app:helicity}. In this work, the mass of $u$ and $d$ quarks are taken to be the dynamical quark mass, $m=5 {\;\rm MeV}$.

\begin{table*}[htb]
\begin{ruledtabular}
\centering
\caption{\label{HAppho1}Helicity amplitudes for the $\gamma^* p \to N_{q^4 \bar q}$ process (through $\gamma^*\to q\bar q$ mechanism) in all configurations of ground state pentaquarks with isospin $I=1/2$, expressed in terms of transition matrix elements $T_{s's}^\lambda$.}
\begin{tabular}{lccccc}
 $q^4\bar q$ configurations & $J^{P}$ &  $M(q^{4}\bar q)$  ({\rm MeV}) & $A_{1/2}^{q^{4}\overline q}$ & $A_{3/2}^{q^{4}\overline q}$ & $S_{1/2}^{q^{4}\overline q}$  
\\
\hline
 $\Psi^{csf}_{[211]_{C}[31]_{FS}[31]_{F}[4]_{S}}(q^{4}\bar q)$ & $\frac{5}{2}^{-}$
 & 2269 & 0 & 0 & 0 
\\
  $\phantom{-}$ & $\frac{3}{2}^{-}$
 & 2025 & 0 & 0 & 0 
\\
 $\Psi^{csf}_{[211]_{C}[31]_{FS}[4]_{F}[31]_{S}}(q^{4}\bar q)$ & $\frac{1}{2}^{-}$ 
 & 2562  & 0 & $\phantom{-}$ & 0 
    \\
 $\phantom{-}$ & $\frac{3}{2}^{-}$   &  2269 & 0 & 0 & 0 
\\
 $\Psi^{csf}_{[211]_{C}[31]_{FS}[31]_{F}[31]_{S}}(q^{4}\bar q)$ & $\frac{1}{2}^{-}$
   & 2123 & $\frac{2}{{3\sqrt{3}}} T^+_{\uparrow\uparrow}$ & $\phantom{-}$ & $\frac{2}{{3\sqrt{3}}} \left(T^0_{\downarrow\uparrow}-2 T^0_{\uparrow\downarrow} \right)$ 
   \\
 $\phantom{-}$ & $\frac{3}{2}^{-}$   &  2049 & $-\frac{4}{{3\sqrt{6}}} T^+_{\uparrow\uparrow}$ & $-\frac{4}{{3\sqrt{2}}} T^+_{\uparrow\uparrow}$ & $-\frac{4}{{3\sqrt{6}}} \left(T^0_{\downarrow\uparrow}+ T^0_{\uparrow\downarrow} \right)$ 
\\
 $\Psi^{csf}_{[211]_{C}[31]_{FS}[31]_{F}[22]_{S}}(q^{4}\bar q)$ & $\frac{1}{2}^{-}$   & 2025 & $-\frac{2}{{\sqrt{3}}} T^+_{\uparrow\uparrow}$ & $\phantom{-}$ & $\frac{2}{{\sqrt{3}}} T^0_{\downarrow\uparrow}$ 
\\
 $\Psi^{csf}_{[211]_{C}[31]_{FS}[22]_{F}[31]_{S}}(q^{4}\bar q)$ & $\frac{1}{2}^{-}$  & 1683 & $-\frac{2}{{3\sqrt{3}}} T^+_{\uparrow\uparrow}$ & $\phantom{-}$ & $-\frac{2}{{3\sqrt{3}}} \left(T^0_{\downarrow\uparrow}-2 T^0_{\uparrow\downarrow} \right)$  \\
  $\phantom{-}$ & $\frac{3}{2}^{-}$  &  2049 & $\frac{4}{{3\sqrt{6}}} T^+_{\uparrow\uparrow}$ & $\frac{4}{{3\sqrt{2}}} T^+_{\uparrow\uparrow}$ & $\frac{4}{{3\sqrt{6}}} \left(T^0_{\downarrow\uparrow}+ T^0_{\uparrow\downarrow} \right)$  \\
\end{tabular}
\end{ruledtabular}
\end{table*}

The helicity amplitudes of pentaquark states, defined in Eq. (\ref{eq:helicity5q}), are calculated for all $q^{4}\bar q$ configurations listed in Table \ref{tab:d1}. The helicity amplitudes for both $J=\frac{1}{2}$ and $\frac{3}{2}$ pentaquark states took the general form, 



\begin{align} \label{generalhe}
A^{(5q)}_{i}
&=\dfrac{1}{\sqrt{2K}}\bra{\Psi_{N^*}^X}\bra{\Psi_{N^*}^{SF}}\bra{\Psi_{N^*}^C} T^{+}(q_1q_2q_3 \gamma^* \rightarrow q_1'q_2'q_3'q'_4\bar {q'_5}) \nonumber\\ 
& \ket{\Psi_{N}^C}\ket{\Psi_{N}^{SF}}\ket{\Psi_{N}^X}\nonumber\\
&=\dfrac{4e}{\sqrt{2K}}\bra{\Psi_{N^*}^X}\bra{\Psi_{N^*}^{SF}}\bra{\Psi_{N^*}^C} T^+_{s' s}\ket{\Psi_{N}^C}\ket{\Psi_{N}^{SF}}\ket{\Psi_{N}^X}\nonumber\\
S^{(5q)}_{1/2}
&=\dfrac{1}{\sqrt{2K}}\bra{\Psi_{N^*}^X}\bra{\Psi_{N^*}^{SF}}\bra{\Psi_{N^*}^C} T^{0}(q_1q_2q_3 \gamma^* \rightarrow q_1'q_2'q_3'q'_4\bar {q'_5}) \nonumber\\ 
& \ket{\Psi_{N}^C}\ket{\Psi_{N}^{SF}}\ket{\Psi_{N}^X}\nonumber\\
&=\dfrac{4e}{\sqrt{2K}}\bra{\Psi_{N^*}^X}\bra{\Psi_{N^*}^{SF}}\bra{\Psi_{N^*}^C} T^0_{s' s}\ket{\Psi_{N}^C}\ket{\Psi_{N}^{SF}}\ket{\Psi_{N}^X}
\end{align}
where $e=\sqrt{4\pi \alpha}$ with $\alpha=1/137$ being the fine-structure constant, and the wave functions of $L=0$ spatial symmetric pentaquark components and proton are written in terms of the spatial, color, spin-flavor parts $\Psi_{N^*}^X$, $\Psi_{N^*}^C$, $\Psi_{N^*}^{SF}$ and $\Psi_N^X$, $\Psi_N^C$, $\Psi_N^{SF}$, respectively. After calculating the matrix elements of color-spin-flavor parts $\bra{\Psi_{N^*}^{SF}}\bra{\Psi_{N^*}^C} T^\lambda_{s' s}\ket{\Psi_{N}^C}\ket{\Psi_{N}^{SF}}$ for all $q^{4}\bar q$ pentaquark configurations, the transition amplitudes, $T^\lambda$ are derived as listed in Table \ref{HAppho1}.

\section{\label{sec4}Results of helicity amplitudes of $N(1520)$ and $N(1535)$}
In the present calculation, we consider the simplest case that the $L=0$ $q^3$ components mix with only one pentaquark state. The pentaquark state in the configuration $\Psi^{csf}_{[211]_{C}[31]_{FS}[22]_{F}[31]_{S}}(q^{4}\bar q)$ is chosen as the representative since it has the lowest mass for $J^P=1/2^-$ and does not mix with other configurations. The helicity amplitudes of other $(q^{4}\bar q)$ configurations differ only by an overall sign or multiplicative factor. For $J= 3/2$ $(q^{4}\bar q)$ which contributes to $\gamma^* p\to N(1520)$ process, we have,
\begin{align} \label{1520heli}
A^{(5q)}_{1/2}
&=\dfrac{e}{\sqrt{2K}}\bra{\Psi_{N^*}^X}\dfrac{4}{3\sqrt{6}} T^+_{\uparrow\uparrow}\ket{\Psi_N^X}  \nonumber\\
A^{(5q)}_{3/2}
&=\dfrac{e}{\sqrt{2K}}\bra{\Psi_{N^*}^X}\dfrac{4}{3\sqrt{2}} T^+_{\uparrow\uparrow}\ket{\Psi_N^X}  \nonumber\\
S^{(5q)}_{1/2}
&=\dfrac{e}{\sqrt{2K}}\bra{\Psi_{N^*}^X}\dfrac{4}{3\sqrt{6}} (T^0_{\uparrow\downarrow}+T^0_{\downarrow\uparrow})\ket{\Psi_N^X} 
\end{align}

For $J= 1/2$ $(q^{4}\bar q)$ which contributes to $\gamma^* p\to N(1535)$ process, the helicity amplitudes take the form,
\begin{align}\label{1535heli}
A^{(5q)}_{1/2}
&=\dfrac{e}{\sqrt{2K}}\bra{\Psi_{N^*}^X}\dfrac{-4}{6\sqrt{3}} T^+_{\uparrow\uparrow}\ket{\Psi_N^X}  \nonumber\\
S^{(5q)}_{1/2}
&=\dfrac{e}{\sqrt{2K}}\bra{\Psi_{N^*}^X}\dfrac{-4}{6\sqrt{3}} (T^0_{\downarrow\uparrow}-2T^0_{\uparrow\downarrow})\ket{\Psi_N^X} 
\end{align}
The helicity amplitudes $A_{1/2}$, $A_{3/2}$, and $S_{1/2}$ for $N(1520)$, as well as $A_{1/2}$ and $S_{1/2}$ for $N(1535)$ are computed for the $L=1$ $q^3$ and $L=0$ $q^4\bar q$ components by integrating over all internal quark momenta of the proton and $N(1520)$ as well as $N(1535)$. 

\begin{figure}[hbt]
\begin{center}
\includegraphics[width=0.45\textwidth]{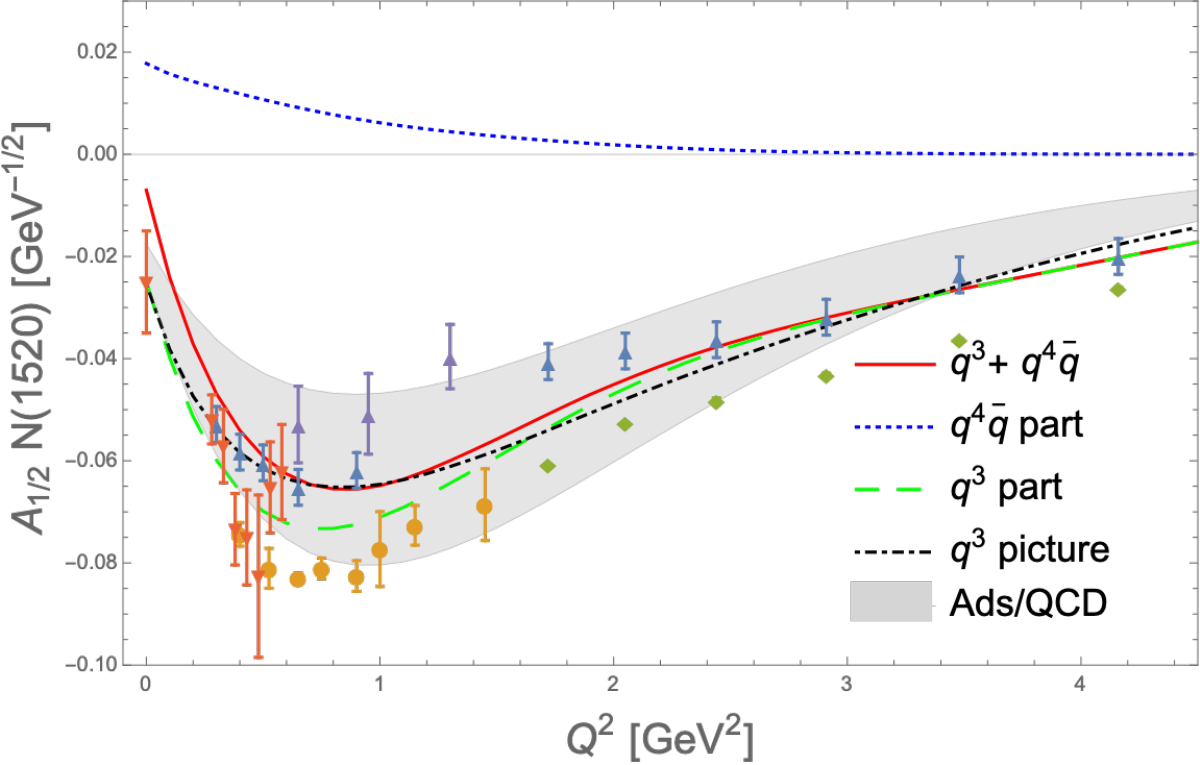}
\includegraphics[width=0.45\textwidth]{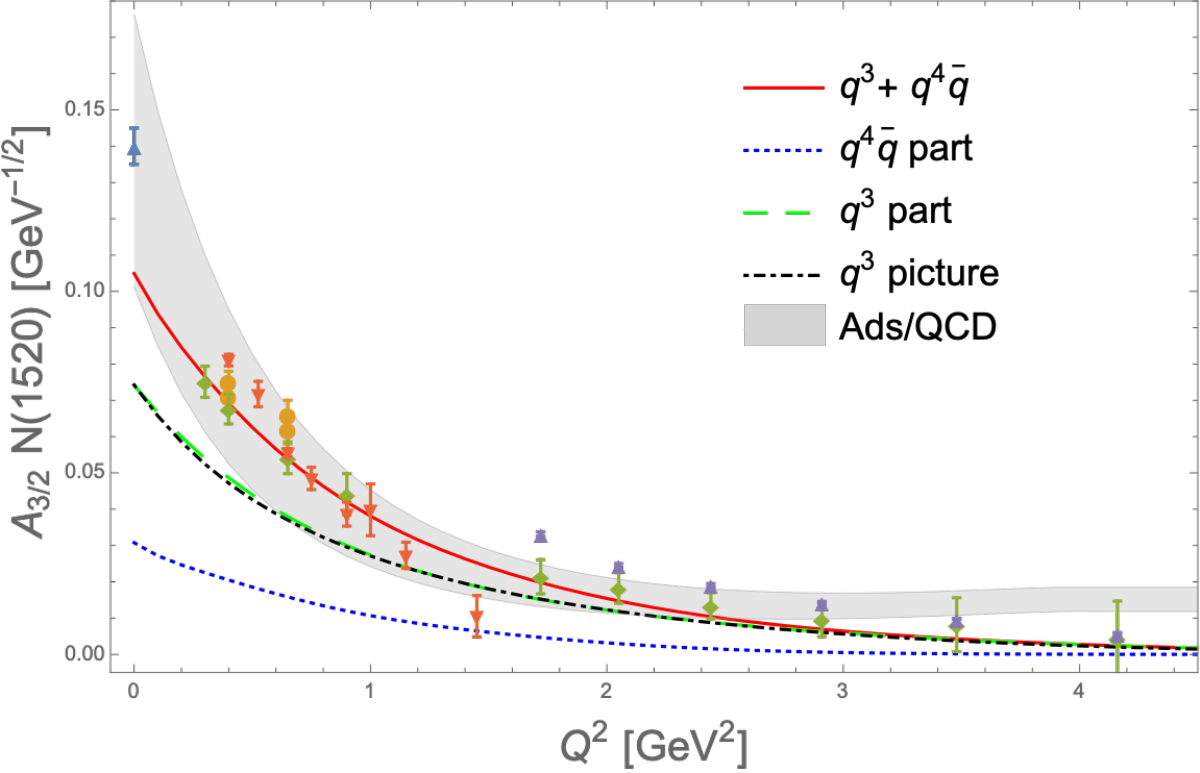}
\includegraphics[width=0.45\textwidth]{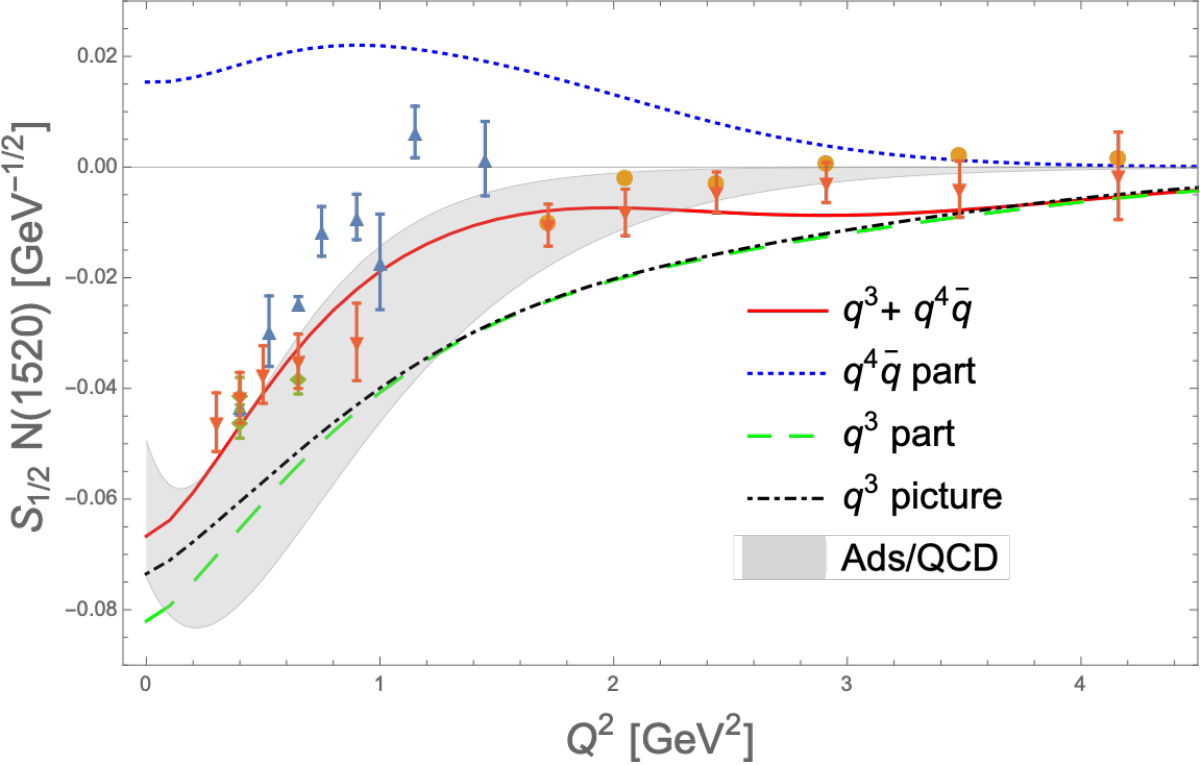}
\end{center}	
\caption{Helicity amplitudes of $N(1520)$ in the mixing picture of three-quark and pentaquark components. Experimental data are taken from CLAS(JLab) data: circles (empty) \cite{PhysRevC.71.015201}, circles (bold) \cite{Aznauryan:2009mx}, triangle (bold) \cite{PhysRevC.86.035203}, triangle (empty) \cite{PhysRevC.93.025206}; MAID(Mainz) data: diamond (empty) \cite{Drechsel:2007if}, diamond (bold) \cite{Tiator:2009mt} and Particle Data Group (PDG) squares (empty) \cite{PDG2024}.}\label{N1520mix3s}
\end{figure}

\begin{figure}[hbt]
\begin{center}
\includegraphics[width=0.45\textwidth]{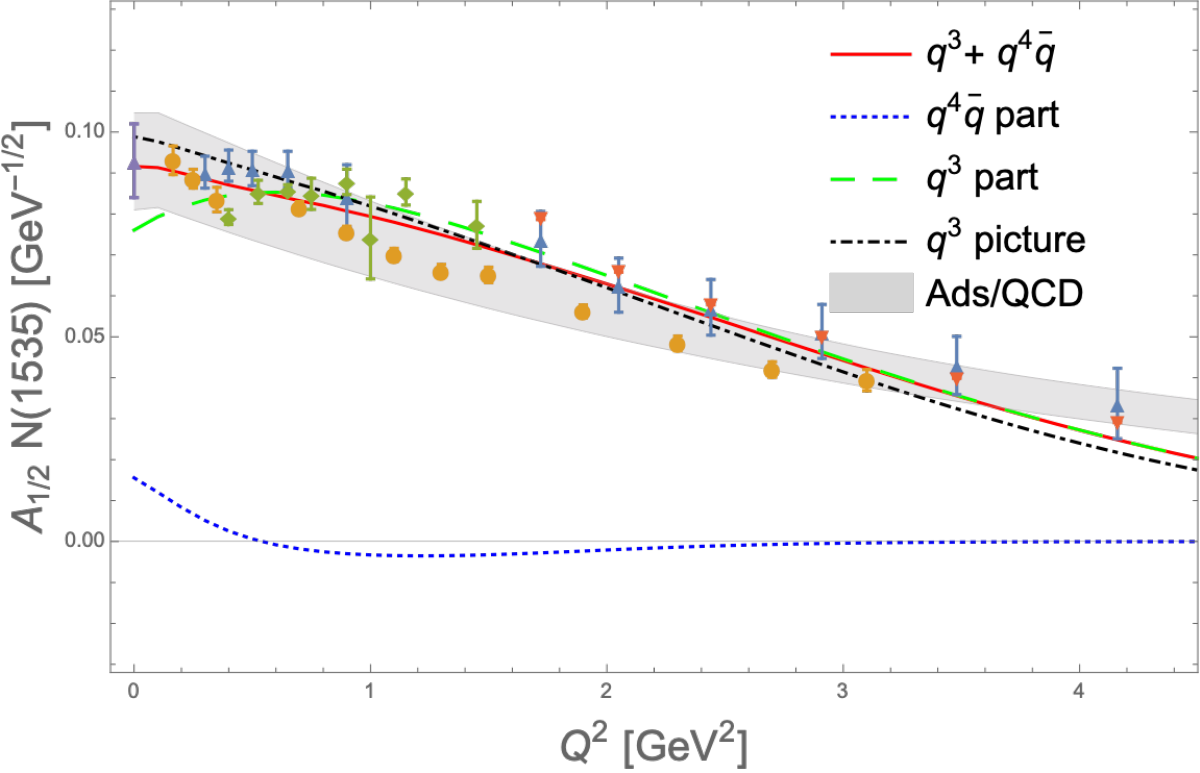}
\includegraphics[width=0.45\textwidth]{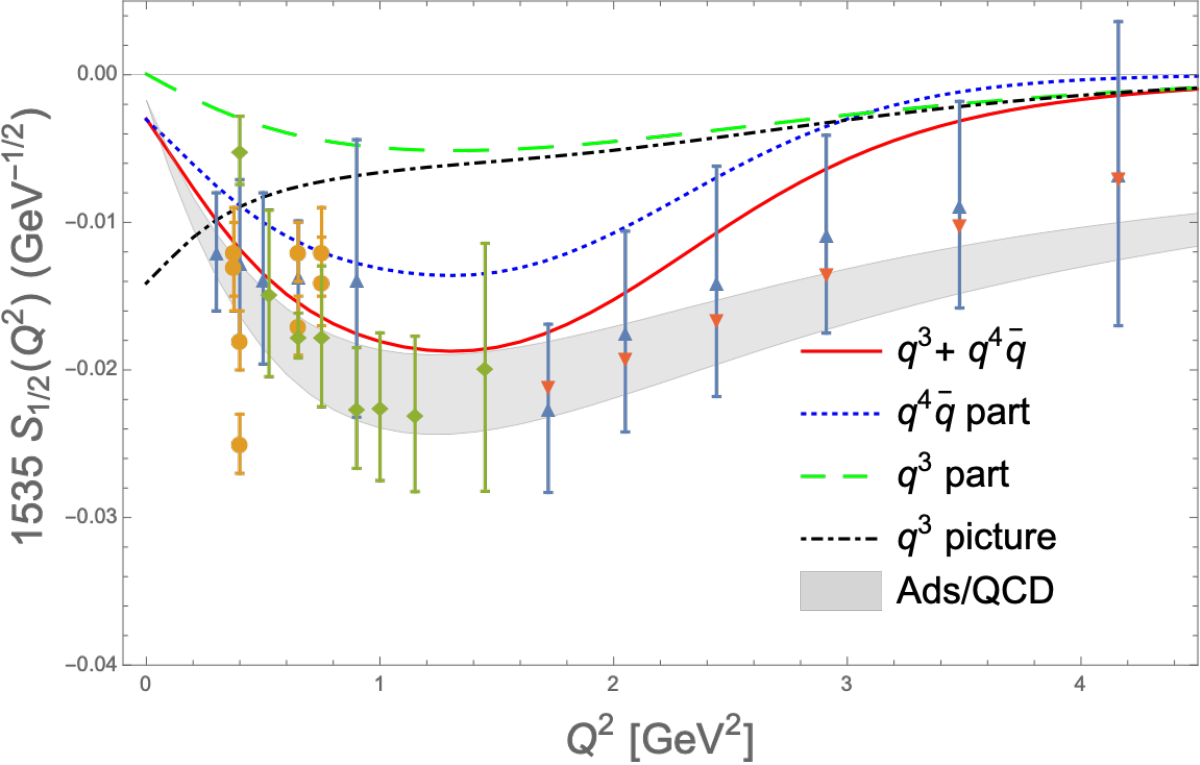}
\end{center}	
		\caption{Helicity amplitudes of $N(1535)$ in the mixing picture of three-quark and pentaquark components. Experimental data are taken from CLAS(JLab) data: circles (empty) \cite{PhysRevC.71.015201}, circles (bold) \cite{Aznauryan:2009mx}, squares (bold) \cite{PhysRevC.76.015204}; MAID(Mainz) data: diamond (empty) \cite{Drechsel:2007if}, diamond (bold) \cite{Tiator:2009mt} and Particle Data Group (PDG) squares (empty) \cite{PDG2024}.}\label{N1535mix3s}
\end{figure}


The theoretical helicity amplitudes $A_{1/2}$, $A_{3/2}$, and $S_{1/2}$ for $N(1520)$, and $A_{1/2}$ and $S_{1/2}$ for $N(1535)$, calculated in the  framework of the mixed three-quark and pentaquark configurations, are fitted to experimental data. In the calculations, the proton spatial wave function is taken from Ref. \cite{PhysRevD.105.016008}, determined by fitting to experimental data of the proton electric form factor, and the spatial wave functions of the $L=1$ $q^3$ and $L=0$ $q^4\bar q$ components of $N(1520)$ and $N(1535)$ as well as the mixing coefficients in Eq. (\ref{eq:ngam}) are adjusted by the fitting. The results of the helicity amplitudes for the $N(1520)$ and $N(1535)$ resonances are shown in Figs. \ref{N1520mix3s} and \ref{N1535mix3s}, where the red solid lines represent the total helicity amplitudes obtained from the mixing picture, the blue dashed lines denote the contributions from the pentaquark component, and green long-dashed lines correspond to the three-quark contributions. For comparison, we include two additional theoretical results for the helicity amplitudes. The first, derived within the same framework as the present work but restricted to pure $L=1$ three-quark configurations \cite{MosEpja}, is shown as black dashed-dotted curves. The second, obtained from soft-wall AdS/QCD approaches \cite{PhysRevD.101.034026,PhysRevD.102.094008}, is represented by shaded bands, that account for variation of model parameters by up to $20\%$. The predictions of the present model exhibit overall consistency with the soft-wall AdS/QCD predictions for both the $N(1520)$ and $N(1535)$ resonances, particularly in the low-$Q^2$ region. 

 The coefficients, $B_{3q}^2$ and $B_{5q}^2$ in Eq. (\ref{eq:ngam}), which represent the contributions of the $q^3$ and $q^4\bar q$ components, are derived,
\begin{align}\label{properties35}
	N(1520): B_{3q}^2: B_{5q}^2 = 0.91: 0.09, \nonumber\\
	N(1535): B_{3q}^2: B_{5q}^2 = 0.75: 0.25
\end{align}

The studies in Refs. \cite{An2009,ZOU2010199}, employing a non-relativistic framework and incorporating pentaquark components in both proton and the $N(1535)$ resonance, achieve an improved description of the helicity amplitude $A_{1/2}$ for $N(1535)$ when assuming pentaquark admixtures of approximately $45\%$ in $N(1535)$ and $20\%$ in the proton. In contrast, our results indicate that a smaller pentaquark component in $N(1535)$ is sufficient to achieve a good agreement with the data. Moreover, all helicity amplitudes of the $N(1520)$ and $N(1535)$ are simultaneously better reproduced in a smaller pentaquark components, suggesting a more moderate role of five-quark configurations.   

As shown in Figs. \ref{N1520mix3s} and \ref{N1535mix3s}, the three-quark contributions in the mixing picture remain largely consistent with 
the ones in the three-quark picture and still roughly describe the experimental data of the helicity amplitudes. It demonstrates that the three-quark $L=1$ states are still the main components in the $N(1520)$ and $N(1535)$ resonances. Notable deviations arise in the $A_{1/2}$ and $S_{1/2}$ amplitudes of $N(1535)$, particularly in the low-$Q^2$ region, where the pentaquark components in $N^*$ yield significant contributions. The pentaquark components in $N(1520)$ contribute constructively to the helicity amplitude $A_{3/2}$ and destructively to $S_{1/2}$. For $N(1535)$ the pentaquark components give a constructive contribution to $S_{1/2}$ and significantly improve the agreement with experimental data, especially for $Q^2 \leq 2\;{\rm GeV^2}$.

\section{\label{sec5} Summary and Discussion}

In the present work we have fitted the helicity amplitudes of the $N(1520)$ and $N(1535)$ in the $\gamma^* p \to N^*$ transition process to experimental data, where the $N(1520)$ and $N(1535)$ are assumed including both the $L=1$ three-quark and $L=0$ symmetric pentaquark component. By introducing the pentaquark component of the $\Psi^{csf}_{[211]_{C}[31]_{FS}[22]_{F}[31]_{S}}(q^{4}\bar q)$ configuration into the $N^*$, a fairly well description of the helicity amplitudes $A_{1/2}$, $A_{3/2}$, and $S_{1/2}$ for the $N(1520)$ and $N(1535)$ resonances over the region $0\leq Q^2\leq 4.5 \;{\rm GeV^2}$ is achieved. The extracted pentaquark components are approximately $9\%$ for $N(1520)$, and $25\%$ for $N(1535)$. 
The clear improvements to the $N(1520)$ and $N(1535)$ helicity amplitudes in the mixing picture suggest that pentaquark components in $N^*$ via the $\gamma^*\to q\bar q$ process may play an important role in the lower $Q^2$ regions, especially for the $S_{1/2}$ amplitudes. 

 
The only notable discrepancies with data occur in the transverse helicity amplitudes of $N(1520)$ at $Q^2 \leq 0.5\;{\rm GeV^2}$ and the longitudinal amplitude of $N(1535)$ at $Q^2 > 2$$\;{\rm GeV^2}$. These deviations may originate from meson cloud contributions, higher-order gluonic exchange contributions, or other dynamical mechanisms, which are not included in the present model. This work serves a starting point for future comprehensive studies that will incorporate contributions of pentaquark components, meson-clouds, and higher-order gluonic exchanges.


\begin{acknowledgments}
This research has received funding support from the NSRF via the Program Management Unit for Human Resources and Institutional Development, Research and Innovation (PMU-B) [grant number B05F640055]. K. X. is additionally supported by SUT (Full-Time66/02/2024) and the NSRF via PMU-B (grant number B13F660067). A. L. and Y. Y. acknowledge the support of (i) Suranaree University of Technology (SUT), (ii) Thailand Science Research and Innovation (TSRI), and (iii) the National Science, Research and Innovation Fund (NSRF), project No. 204211.
\end{acknowledgments}

\appendix\normalsize

\section{Matrix elements of $\gamma^* \to q \bar q$ process}\label{app:helicity}

The matrix elements $T^{\lambda}_{s's}$ in Eq. (\ref{transitionapp}) for the process $\gamma^*\to q\bar q$ with the helicity $\lambda=0,1$ are derived in the standard language of quantum field theory,
\begin{eqnarray}
\label{matrixelement1}
&T^0_{\uparrow\uparrow}=\sqrt{\dfrac{(E'+m)(E+m)}{4E'E}}\left[ \dfrac{p_-}{E+m}+\dfrac{p'_-}{E'+m}\right], \nonumber\\
&T^0_{\uparrow\downarrow}= -\sqrt{\dfrac{(E'+m)(E+m)}{4E'E}}\left[ \dfrac{p_z}{E+m}+\dfrac{p'_z}{E'+m}\right], \nonumber\\
&T^0_{\downarrow\uparrow}=-\sqrt{\dfrac{(E'+m)(E+m)}{4E'E}}\left[ \dfrac{p_z}{E+m}+\dfrac{p'_z}{E'+m}\right], \nonumber\\
&T^0_{\downarrow\downarrow}=-\sqrt{\dfrac{(E'+m)(E+m)}{4E'E}}\left[ \dfrac{p_+}{E+m}+\dfrac{p'_+}{E'+m}\right], \nonumber\\
&T^+_{\uparrow\uparrow}=\sqrt{2}\sqrt{\dfrac{(E'+m)(E+m)}{4E'E}}\left[ 1-\dfrac{p'_z p_z}{(E'+m)(E+m)}\right], \nonumber\\
&T^+_{\uparrow\downarrow}=-\sqrt{2}\sqrt{\dfrac{(E'+m)(E+m)}{4E'E}}\left[ \dfrac{\sqrt{2}p'_z p_+}{(E'+m)(E+m)}\right], \nonumber\\
&T^+_{\downarrow\uparrow}=-\sqrt{2}\sqrt{\dfrac{(E'+m)(E+m)}{4E'E}}\left[ \dfrac{\sqrt{2}p'_+p_z}{(E'+m)(E+m)}\right],\nonumber\\
&T^+_{\downarrow\downarrow}=-\sqrt{2}\sqrt{\dfrac{(E'+m)(E+m)}{4E'E}}\left[\dfrac{2p'_+ p_+}{(E'+m)(E+m)}\right].
\end{eqnarray} 
where $E'$($p'$) and $E$($p$) are respectively the energies (momenta) of the quark and antiquark, and $p_{\pm}=\frac{1}{\sqrt{2}}(p_{x}\pm ip_{y})$. The quark and antiquark Dirac spinors take the form,
\begin{eqnarray}
u_{s'} (p')
&= &\sqrt{\dfrac{E'+m}{2m}}\left(
\begin{array}{c}
\chi_{s'}  \\
\dfrac{\vec \sigma\cdot\vec p' }{E'+m} \chi_{s'}
\end{array}
\right),\; \nonumber\\
v_{s} (p)
&=& \sqrt{\dfrac{E+m}{2m}} \left(
\begin{array}{c}
\dfrac{\vec \sigma\cdot\vec p}{E+m} \bar \chi_{s}  \\
\bar \chi_{s}   
\end{array}
\right).\;
\end{eqnarray} 
where the spin states of the antiquark are linked to the quark spin sates by the charge conjugation $C=i \gamma^2\gamma^0$, that is, $\bar \chi_{s}=-i \sigma_2 \chi_{s}$. Thus we have,
\begin{eqnarray}
 \uparrow
= \left(
\begin{array}{c}
1  \\
0   
\end{array}
\right) ,\;
\downarrow
= \left(
\begin{array}{c}
0  \\
1   
\end{array}
\right) ,\;
 \bar{ \uparrow}
= \left(
\begin{array}{c}
0 \\
1
\end{array}
\right),\;
 \bar{\downarrow}
= \left(
\begin{array}{c}
-1 \\
0
\end{array}
\right)
\nonumber \\
\end{eqnarray}

\bibliography{helicityref}

\begin{thebibliography}{43}%
\makeatletter
\providecommand \@ifxundefined [1]{%
 \@ifx{#1\undefined}
}%
\providecommand \@ifnum [1]{%
 \ifnum #1\expandafter \@firstoftwo
 \else \expandafter \@secondoftwo
 \fi
}%
\providecommand \@ifx [1]{%
 \ifx #1\expandafter \@firstoftwo
 \else \expandafter \@secondoftwo
 \fi
}%
\providecommand \natexlab [1]{#1}%
\providecommand \enquote  [1]{``#1''}%
\providecommand \bibnamefont  [1]{#1}%
\providecommand \bibfnamefont [1]{#1}%
\providecommand \citenamefont [1]{#1}%
\providecommand \href@noop [0]{\@secondoftwo}%
\providecommand \href [0]{\begingroup \@sanitize@url \@href}%
\providecommand \@href[1]{\@@startlink{#1}\@@href}%
\providecommand \@@href[1]{\endgroup#1\@@endlink}%
\providecommand \@sanitize@url [0]{\catcode `\\12\catcode `\$12\catcode
  `\&12\catcode `\#12\catcode `\^12\catcode `\_12\catcode `\%12\relax}%
\providecommand \@@startlink[1]{}%
\providecommand \@@endlink[0]{}%
\providecommand \url  [0]{\begingroup\@sanitize@url \@url }%
\providecommand \@url [1]{\endgroup\@href {#1}{\urlprefix }}%
\providecommand \urlprefix  [0]{URL }%
\providecommand \Eprint [0]{\href }%
\providecommand \doibase [0]{https://doi.org/}%
\providecommand \selectlanguage [0]{\@gobble}%
\providecommand \bibinfo  [0]{\@secondoftwo}%
\providecommand \bibfield  [0]{\@secondoftwo}%
\providecommand \translation [1]{[#1]}%
\providecommand \BibitemOpen [0]{}%
\providecommand \bibitemStop [0]{}%
\providecommand \bibitemNoStop [0]{.\EOS\space}%
\providecommand \EOS [0]{\spacefactor3000\relax}%
\providecommand \BibitemShut  [1]{\csname bibitem#1\endcsname}%
\let\auto@bib@innerbib\@empty
\bibitem [{\citenamefont {Thompson}\ \emph {et~al.}(2001)\citenamefont
  {Thompson} \emph {et~al.}}]{PhysRevLett.86.1702}%
  \BibitemOpen
  \bibfield  {author} {\bibinfo {author} {\bibfnamefont {R.}~\bibnamefont
  {Thompson}} \emph {et~al.} (\bibinfo {collaboration} {The CLAS
  Collaboration}),\ }\href {https://doi.org/10.1103/PhysRevLett.86.1702}
  {\bibfield  {journal} {\bibinfo  {journal} {Phys. Rev. Lett.}\ }\textbf
  {\bibinfo {volume} {86}},\ \bibinfo {pages} {1702} (\bibinfo {year}
  {2001})}\BibitemShut {NoStop}%
\bibitem [{\citenamefont {Aznauryan}(2003)}]{PhysRevC.68.065204}%
  \BibitemOpen
  \bibfield  {author} {\bibinfo {author} {\bibfnamefont {I.~G.}\ \bibnamefont
  {Aznauryan}},\ }\href {https://doi.org/10.1103/PhysRevC.68.065204} {\bibfield
   {journal} {\bibinfo  {journal} {Phys. Rev. C}\ }\textbf {\bibinfo {volume}
  {68}},\ \bibinfo {pages} {065204} (\bibinfo {year} {2003})}\BibitemShut
  {NoStop}%
\bibitem [{\citenamefont {Dalton}\ \emph {et~al.}(2009)\citenamefont {Dalton}
  \emph {et~al.}}]{PhysRevC.80.015205}%
  \BibitemOpen
  \bibfield  {author} {\bibinfo {author} {\bibfnamefont {M.~M.}\ \bibnamefont
  {Dalton}} \emph {et~al.},\ }\href
  {https://doi.org/10.1103/PhysRevC.80.015205} {\bibfield  {journal} {\bibinfo
  {journal} {Phys. Rev. C}\ }\textbf {\bibinfo {volume} {80}},\ \bibinfo
  {pages} {015205} (\bibinfo {year} {2009})}\BibitemShut {NoStop}%
\bibitem [{\citenamefont {Aznauryan}\ \emph {et~al.}(2008)\citenamefont
  {Aznauryan}, \citenamefont {Burkert}, \citenamefont {Kim}, \citenamefont
  {Park}, \citenamefont {Adams}, \citenamefont {Amaryan}, \citenamefont
  {Ambrozewicz}, \citenamefont {Anghinolfi}, \citenamefont {Asryan},
  \citenamefont {Avakian}, \citenamefont {Bagdasaryan}, \citenamefont
  {Baillie}, \citenamefont {Ball}, \citenamefont {Baltzell}, \citenamefont
  {Barrow} \emph {et~al.}}]{PhysRevC.78.045209}%
  \BibitemOpen
  \bibfield  {author} {\bibinfo {author} {\bibfnamefont {I.~G.}\ \bibnamefont
  {Aznauryan}}, \bibinfo {author} {\bibfnamefont {V.~D.}\ \bibnamefont
  {Burkert}}, \bibinfo {author} {\bibfnamefont {W.}~\bibnamefont {Kim}},
  \bibinfo {author} {\bibfnamefont {K.}~\bibnamefont {Park}}, \bibinfo {author}
  {\bibfnamefont {G.}~\bibnamefont {Adams}}, \bibinfo {author} {\bibfnamefont
  {M.~J.}\ \bibnamefont {Amaryan}}, \bibinfo {author} {\bibfnamefont
  {P.}~\bibnamefont {Ambrozewicz}}, \bibinfo {author} {\bibfnamefont
  {M.}~\bibnamefont {Anghinolfi}}, \bibinfo {author} {\bibfnamefont
  {G.}~\bibnamefont {Asryan}}, \bibinfo {author} {\bibfnamefont
  {H.}~\bibnamefont {Avakian}}, \bibinfo {author} {\bibfnamefont
  {H.}~\bibnamefont {Bagdasaryan}}, \bibinfo {author} {\bibfnamefont
  {N.}~\bibnamefont {Baillie}}, \bibinfo {author} {\bibfnamefont {J.~P.}\
  \bibnamefont {Ball}}, \bibinfo {author} {\bibfnamefont {N.~A.}\ \bibnamefont
  {Baltzell}}, \bibinfo {author} {\bibfnamefont {S.}~\bibnamefont {Barrow}},
  \emph {et~al.} (\bibinfo {collaboration} {CLAS Collaboration}),\ }\href
  {https://doi.org/10.1103/PhysRevC.78.045209} {\bibfield  {journal} {\bibinfo
  {journal} {Phys. Rev. C}\ }\textbf {\bibinfo {volume} {78}},\ \bibinfo
  {pages} {045209} (\bibinfo {year} {2008})}\BibitemShut {NoStop}%
\bibitem [{\citenamefont {Aznauryan}\ \emph {et~al.}(2005)\citenamefont
  {Aznauryan} \emph {et~al.}}]{PhysRevC.71.015201}%
  \BibitemOpen
  \bibfield  {author} {\bibinfo {author} {\bibfnamefont {I.~G.}\ \bibnamefont
  {Aznauryan}} \emph {et~al.},\ }\href
  {https://doi.org/10.1103/PhysRevC.71.015201} {\bibfield  {journal} {\bibinfo
  {journal} {Phys. Rev. C}\ }\textbf {\bibinfo {volume} {71}},\ \bibinfo
  {pages} {015201} (\bibinfo {year} {2005})}\BibitemShut {NoStop}%
\bibitem [{\citenamefont {Aznauryan}\ \emph {et~al.}(2009)\citenamefont
  {Aznauryan} \emph {et~al.}}]{Aznauryan:2009mx}%
  \BibitemOpen
  \bibfield  {author} {\bibinfo {author} {\bibfnamefont {I.~G.}\ \bibnamefont
  {Aznauryan}} \emph {et~al.} (\bibinfo {collaboration} {CLAS}),\ }\href
  {https://doi.org/10.1103/PhysRevC.80.055203} {\bibfield  {journal} {\bibinfo
  {journal} {Phys. Rev. C}\ }\textbf {\bibinfo {volume} {80}},\ \bibinfo
  {pages} {055203} (\bibinfo {year} {2009})},\ \Eprint
  {https://arxiv.org/abs/0909.2349} {arXiv:0909.2349 [nucl-ex]} \BibitemShut
  {NoStop}%
\bibitem [{\citenamefont {Denizli}\ \emph {et~al.}(2007)\citenamefont {Denizli}
  \emph {et~al.}}]{PhysRevC.76.015204}%
  \BibitemOpen
  \bibfield  {author} {\bibinfo {author} {\bibfnamefont {H.}~\bibnamefont
  {Denizli}} \emph {et~al.} (\bibinfo {collaboration} {CLAS Collaboration}),\
  }\href {https://doi.org/10.1103/PhysRevC.76.015204} {\bibfield  {journal}
  {\bibinfo  {journal} {Phys. Rev. C}\ }\textbf {\bibinfo {volume} {76}},\
  \bibinfo {pages} {015204} (\bibinfo {year} {2007})}\BibitemShut {NoStop}%
\bibitem [{\citenamefont {Mokeev}\ \emph {et~al.}(2012)\citenamefont {Mokeev}
  \emph {et~al.}}]{PhysRevC.86.035203}%
  \BibitemOpen
  \bibfield  {author} {\bibinfo {author} {\bibfnamefont {V.~I.}\ \bibnamefont
  {Mokeev}} \emph {et~al.} (\bibinfo {collaboration} {CLAS}),\ }\href
  {https://doi.org/10.1103/PhysRevC.86.035203} {\bibfield  {journal} {\bibinfo
  {journal} {Phys. Rev. C}\ }\textbf {\bibinfo {volume} {86}},\ \bibinfo
  {pages} {035203} (\bibinfo {year} {2012})},\ \Eprint
  {https://arxiv.org/abs/1205.3948} {arXiv:1205.3948 [nucl-ex]} \BibitemShut
  {NoStop}%
\bibitem [{\citenamefont {Mokeev}\ \emph {et~al.}(2016)\citenamefont {Mokeev},
  \citenamefont {Burkert}, \citenamefont {Carman}, \citenamefont {Elouadrhiri},
  \citenamefont {Fedotov}, \citenamefont {Golovatch}, \citenamefont {Gothe},
  \citenamefont {Hicks}, \citenamefont {Ishkhanov}, \citenamefont {Isupov},\
  and\ \citenamefont {Skorodumina}}]{PhysRevC.93.025206}%
  \BibitemOpen
  \bibfield  {author} {\bibinfo {author} {\bibfnamefont {V.~I.}\ \bibnamefont
  {Mokeev}}, \bibinfo {author} {\bibfnamefont {V.~D.}\ \bibnamefont {Burkert}},
  \bibinfo {author} {\bibfnamefont {D.~S.}\ \bibnamefont {Carman}}, \bibinfo
  {author} {\bibfnamefont {L.}~\bibnamefont {Elouadrhiri}}, \bibinfo {author}
  {\bibfnamefont {G.~V.}\ \bibnamefont {Fedotov}}, \bibinfo {author}
  {\bibfnamefont {E.~N.}\ \bibnamefont {Golovatch}}, \bibinfo {author}
  {\bibfnamefont {R.~W.}\ \bibnamefont {Gothe}}, \bibinfo {author}
  {\bibfnamefont {K.}~\bibnamefont {Hicks}}, \bibinfo {author} {\bibfnamefont
  {B.~S.}\ \bibnamefont {Ishkhanov}}, \bibinfo {author} {\bibfnamefont {E.~L.}\
  \bibnamefont {Isupov}},\ and\ \bibinfo {author} {\bibfnamefont
  {I.}~\bibnamefont {Skorodumina}},\ }\href
  {https://doi.org/10.1103/PhysRevC.93.025206} {\bibfield  {journal} {\bibinfo
  {journal} {Phys. Rev. C}\ }\textbf {\bibinfo {volume} {93}},\ \bibinfo
  {pages} {025206} (\bibinfo {year} {2016})}\BibitemShut {NoStop}%
\bibitem [{\citenamefont {Drechsel}\ \emph
  {et~al.}(2007{\natexlab{a}})\citenamefont {Drechsel}, \citenamefont
  {Kamalov},\ and\ \citenamefont {Tiator}}]{Drechsel:2007if}%
  \BibitemOpen
  \bibfield  {author} {\bibinfo {author} {\bibfnamefont {D.}~\bibnamefont
  {Drechsel}}, \bibinfo {author} {\bibfnamefont {S.~S.}\ \bibnamefont
  {Kamalov}},\ and\ \bibinfo {author} {\bibfnamefont {L.}~\bibnamefont
  {Tiator}},\ }\href {https://doi.org/10.1140/epja/i2007-10490-6} {\bibfield
  {journal} {\bibinfo  {journal} {Eur. Phys. J. A}\ }\textbf {\bibinfo {volume}
  {34}},\ \bibinfo {pages} {69} (\bibinfo {year} {2007}{\natexlab{a}})},\
  \Eprint {https://arxiv.org/abs/0710.0306} {arXiv:0710.0306 [nucl-th]}
  \BibitemShut {NoStop}%
\bibitem [{\citenamefont {Tiator}\ \emph {et~al.}(2009)\citenamefont {Tiator},
  \citenamefont {Drechsel}, \citenamefont {Kamalov},\ and\ \citenamefont
  {Vanderhaeghen}}]{Tiator:2009mt}%
  \BibitemOpen
  \bibfield  {author} {\bibinfo {author} {\bibfnamefont {L.}~\bibnamefont
  {Tiator}}, \bibinfo {author} {\bibfnamefont {D.}~\bibnamefont {Drechsel}},
  \bibinfo {author} {\bibfnamefont {S.~S.}\ \bibnamefont {Kamalov}},\ and\
  \bibinfo {author} {\bibfnamefont {M.}~\bibnamefont {Vanderhaeghen}},\ }\href
  {https://doi.org/10.1088/1674-1137/33/12/005} {\bibfield  {journal} {\bibinfo
   {journal} {Chin. Phys. C}\ }\textbf {\bibinfo {volume} {33}},\ \bibinfo
  {pages} {1069} (\bibinfo {year} {2009})},\ \Eprint
  {https://arxiv.org/abs/0909.2335} {arXiv:0909.2335 [nucl-th]} \BibitemShut
  {NoStop}%
\bibitem [{\citenamefont {Tiator}\ \emph {et~al.}(2011)\citenamefont {Tiator},
  \citenamefont {Drechsel}, \citenamefont {Kamalov},\ and\ \citenamefont
  {Vanderhaeghen}}]{Tiator:2011pw}%
  \BibitemOpen
  \bibfield  {author} {\bibinfo {author} {\bibfnamefont {L.}~\bibnamefont
  {Tiator}}, \bibinfo {author} {\bibfnamefont {D.}~\bibnamefont {Drechsel}},
  \bibinfo {author} {\bibfnamefont {S.~S.}\ \bibnamefont {Kamalov}},\ and\
  \bibinfo {author} {\bibfnamefont {M.}~\bibnamefont {Vanderhaeghen}},\ }\href
  {https://doi.org/10.1140/epjst/e2011-01488-9} {\bibfield  {journal} {\bibinfo
   {journal} {Eur. Phys. J. ST}\ }\textbf {\bibinfo {volume} {198}},\ \bibinfo
  {pages} {141} (\bibinfo {year} {2011})},\ \Eprint
  {https://arxiv.org/abs/1109.6745} {arXiv:1109.6745 [nucl-th]} \BibitemShut
  {NoStop}%
\bibitem [{\citenamefont {Drechsel}\ \emph
  {et~al.}(2007{\natexlab{b}})\citenamefont {Drechsel}, \citenamefont
  {Kamalov},\ and\ \citenamefont {Tiator}}]{Drechsel2007}%
  \BibitemOpen
  \bibfield  {author} {\bibinfo {author} {\bibfnamefont {D.}~\bibnamefont
  {Drechsel}}, \bibinfo {author} {\bibfnamefont {S.~S.}\ \bibnamefont
  {Kamalov}},\ and\ \bibinfo {author} {\bibfnamefont {L.}~\bibnamefont
  {Tiator}},\ }\href {https://doi.org/10.1140/epja/i2007-10490-6} {\bibfield
  {journal} {\bibinfo  {journal} {The European Physical Journal A}\ }\textbf
  {\bibinfo {volume} {34}},\ \bibinfo {pages} {69} (\bibinfo {year}
  {2007}{\natexlab{b}})}\BibitemShut {NoStop}%
\bibitem [{\citenamefont {Aznauryan}\ and\ \citenamefont
  {Burkert}(2012{\natexlab{a}})}]{AZNAURYAN20121}%
  \BibitemOpen
  \bibfield  {author} {\bibinfo {author} {\bibfnamefont {I.}~\bibnamefont
  {Aznauryan}}\ and\ \bibinfo {author} {\bibfnamefont {V.}~\bibnamefont
  {Burkert}},\ }\href
  {https://doi.org/https://doi.org/10.1016/j.ppnp.2011.08.001} {\bibfield
  {journal} {\bibinfo  {journal} {Progress in Particle and Nuclear Physics}\
  }\textbf {\bibinfo {volume} {67}},\ \bibinfo {pages} {1} (\bibinfo {year}
  {2012}{\natexlab{a}})}\BibitemShut {NoStop}%
\bibitem [{\citenamefont {Ramalho}\ and\ \citenamefont
  {Pe\~na}(2024)}]{RAMALHO2024104097}%
  \BibitemOpen
  \bibfield  {author} {\bibinfo {author} {\bibfnamefont {G.}~\bibnamefont
  {Ramalho}}\ and\ \bibinfo {author} {\bibfnamefont {M.~T.}\ \bibnamefont
  {Pe\~na}},\ }\href
  {https://doi.org/https://doi.org/10.1016/j.ppnp.2024.104097} {\bibfield
  {journal} {\bibinfo  {journal} {Progress in Particle and Nuclear Physics}\
  }\textbf {\bibinfo {volume} {136}},\ \bibinfo {pages} {104097} (\bibinfo
  {year} {2024})}\BibitemShut {NoStop}%
\bibitem [{\citenamefont {Burkert}\ \emph {et~al.}(2003)\citenamefont
  {Burkert}, \citenamefont {De~Vita}, \citenamefont {Battaglieri},
  \citenamefont {Ripani},\ and\ \citenamefont {Mokeev}}]{PhysRevC.67.035204}%
  \BibitemOpen
  \bibfield  {author} {\bibinfo {author} {\bibfnamefont {V.~D.}\ \bibnamefont
  {Burkert}}, \bibinfo {author} {\bibfnamefont {R.}~\bibnamefont {De~Vita}},
  \bibinfo {author} {\bibfnamefont {M.}~\bibnamefont {Battaglieri}}, \bibinfo
  {author} {\bibfnamefont {M.}~\bibnamefont {Ripani}},\ and\ \bibinfo {author}
  {\bibfnamefont {V.}~\bibnamefont {Mokeev}},\ }\href
  {https://doi.org/10.1103/PhysRevC.67.035204} {\bibfield  {journal} {\bibinfo
  {journal} {Phys. Rev. C}\ }\textbf {\bibinfo {volume} {67}},\ \bibinfo
  {pages} {035204} (\bibinfo {year} {2003})}\BibitemShut {NoStop}%
\bibitem [{\citenamefont {Capstick}\ and\ \citenamefont
  {Keister}(1995)}]{Capstick:1994ne}%
  \BibitemOpen
  \bibfield  {author} {\bibinfo {author} {\bibfnamefont {S.}~\bibnamefont
  {Capstick}}\ and\ \bibinfo {author} {\bibfnamefont {B.~D.}\ \bibnamefont
  {Keister}},\ }\href {https://doi.org/10.1103/PhysRevD.51.3598} {\bibfield
  {journal} {\bibinfo  {journal} {Phys. Rev. D}\ }\textbf {\bibinfo {volume}
  {51}},\ \bibinfo {pages} {3598} (\bibinfo {year} {1995})},\ \Eprint
  {https://arxiv.org/abs/nucl-th/9411016} {arXiv:nucl-th/9411016} \BibitemShut
  {NoStop}%
\bibitem [{\citenamefont {Pace}\ \emph {et~al.}(2000)\citenamefont {Pace},
  \citenamefont {Salmè}, \citenamefont {Cardarelli},\ and\ \citenamefont
  {Simula}}]{PACE200033}%
  \BibitemOpen
  \bibfield  {author} {\bibinfo {author} {\bibfnamefont {E.}~\bibnamefont
  {Pace}}, \bibinfo {author} {\bibfnamefont {G.}~\bibnamefont {Salmè}},
  \bibinfo {author} {\bibfnamefont {F.}~\bibnamefont {Cardarelli}},\ and\
  \bibinfo {author} {\bibfnamefont {S.}~\bibnamefont {Simula}},\ }\href
  {https://doi.org/https://doi.org/10.1016/S0375-9474(00)00007-5} {\bibfield
  {journal} {\bibinfo  {journal} {Nuclear Physics A}\ }\textbf {\bibinfo
  {volume} {666-667}},\ \bibinfo {pages} {33} (\bibinfo {year} {2000})},\
  \bibinfo {note} {the Structure of the Nucleon}\BibitemShut {NoStop}%
\bibitem [{\citenamefont {Obukhovsky}\ \emph {et~al.}(2019)\citenamefont
  {Obukhovsky}, \citenamefont {Faessler}, \citenamefont {Fedorov},
  \citenamefont {Gutsche},\ and\ \citenamefont
  {Lyubovitskij}}]{PhysRevD.100.094013}%
  \BibitemOpen
  \bibfield  {author} {\bibinfo {author} {\bibfnamefont {I.~T.}\ \bibnamefont
  {Obukhovsky}}, \bibinfo {author} {\bibfnamefont {A.}~\bibnamefont
  {Faessler}}, \bibinfo {author} {\bibfnamefont {D.~K.}\ \bibnamefont
  {Fedorov}}, \bibinfo {author} {\bibfnamefont {T.}~\bibnamefont {Gutsche}},\
  and\ \bibinfo {author} {\bibfnamefont {V.~E.}\ \bibnamefont {Lyubovitskij}},\
  }\href {https://doi.org/10.1103/PhysRevD.100.094013} {\bibfield  {journal}
  {\bibinfo  {journal} {Phys. Rev. D}\ }\textbf {\bibinfo {volume} {100}},\
  \bibinfo {pages} {094013} (\bibinfo {year} {2019})}\BibitemShut {NoStop}%
\bibitem [{\citenamefont {Aznauryan}\ and\ \citenamefont
  {Burkert}(2012{\natexlab{b}})}]{PhysRevC.85.055202}%
  \BibitemOpen
  \bibfield  {author} {\bibinfo {author} {\bibfnamefont {I.~G.}\ \bibnamefont
  {Aznauryan}}\ and\ \bibinfo {author} {\bibfnamefont {V.~D.}\ \bibnamefont
  {Burkert}},\ }\href {https://doi.org/10.1103/PhysRevC.85.055202} {\bibfield
  {journal} {\bibinfo  {journal} {Phys. Rev. C}\ }\textbf {\bibinfo {volume}
  {85}},\ \bibinfo {pages} {055202} (\bibinfo {year}
  {2012}{\natexlab{b}})}\BibitemShut {NoStop}%
\bibitem [{\citenamefont {Aznauryan}\ and\ \citenamefont
  {Burkert}(2017)}]{PhysRevC.95.065207}%
  \BibitemOpen
  \bibfield  {author} {\bibinfo {author} {\bibfnamefont {I.~G.}\ \bibnamefont
  {Aznauryan}}\ and\ \bibinfo {author} {\bibfnamefont {V.~D.}\ \bibnamefont
  {Burkert}},\ }\href {https://doi.org/10.1103/PhysRevC.95.065207} {\bibfield
  {journal} {\bibinfo  {journal} {Phys. Rev. C}\ }\textbf {\bibinfo {volume}
  {95}},\ \bibinfo {pages} {065207} (\bibinfo {year} {2017})}\BibitemShut
  {NoStop}%
\bibitem [{\citenamefont {Giannini}\ and\ \citenamefont
  {Santopinto}(2015)}]{Giannini:2015zia}%
  \BibitemOpen
  \bibfield  {author} {\bibinfo {author} {\bibfnamefont {M.~M.}\ \bibnamefont
  {Giannini}}\ and\ \bibinfo {author} {\bibfnamefont {E.}~\bibnamefont
  {Santopinto}},\ }\href {https://doi.org/10.6122/CJP.20150120} {\bibfield
  {journal} {\bibinfo  {journal} {Chin. J. Phys.}\ }\textbf {\bibinfo {volume}
  {53}},\ \bibinfo {pages} {020301} (\bibinfo {year} {2015})},\ \Eprint
  {https://arxiv.org/abs/1501.03722} {arXiv:1501.03722 [nucl-th]} \BibitemShut
  {NoStop}%
\bibitem [{\citenamefont {Giannini}\ \emph {et~al.}(2001)\citenamefont
  {Giannini}, \citenamefont {Santopinto},\ and\ \citenamefont
  {Vassallo}}]{Giannini:2001kb}%
  \BibitemOpen
  \bibfield  {author} {\bibinfo {author} {\bibfnamefont {M.~M.}\ \bibnamefont
  {Giannini}}, \bibinfo {author} {\bibfnamefont {E.}~\bibnamefont
  {Santopinto}},\ and\ \bibinfo {author} {\bibfnamefont {A.}~\bibnamefont
  {Vassallo}},\ }\href {https://doi.org/10.1007/s10050-001-8668-y} {\bibfield
  {journal} {\bibinfo  {journal} {Eur. Phys. J. A}\ }\textbf {\bibinfo {volume}
  {12}},\ \bibinfo {pages} {447} (\bibinfo {year} {2001})},\ \Eprint
  {https://arxiv.org/abs/nucl-th/0111073} {arXiv:nucl-th/0111073} \BibitemShut
  {NoStop}%
\bibitem [{\citenamefont {Santopinto}\ and\ \citenamefont
  {Giannini}(2012)}]{PhysRevC.86.065202}%
  \BibitemOpen
  \bibfield  {author} {\bibinfo {author} {\bibfnamefont {E.}~\bibnamefont
  {Santopinto}}\ and\ \bibinfo {author} {\bibfnamefont {M.~M.}\ \bibnamefont
  {Giannini}},\ }\href {https://doi.org/10.1103/PhysRevC.86.065202} {\bibfield
  {journal} {\bibinfo  {journal} {Phys. Rev. C}\ }\textbf {\bibinfo {volume}
  {86}},\ \bibinfo {pages} {065202} (\bibinfo {year} {2012})}\BibitemShut
  {NoStop}%
\bibitem [{\citenamefont {Braun}\ \emph {et~al.}(2009)\citenamefont {Braun},
  \citenamefont {G\"ockeler}, \citenamefont {Horsley}, \citenamefont
  {Kaltenbrunner}, \citenamefont {Lenz}, \citenamefont {Nakamura},
  \citenamefont {Pleiter}, \citenamefont {Rakow}, \citenamefont {Rohrwild},
  \citenamefont {Sch\"afer}, \citenamefont {Schierholz}, \citenamefont
  {St\"uben}, \citenamefont {Warkentin},\ and\ \citenamefont
  {Zanotti}}]{PhysRevLett.103.072001}%
  \BibitemOpen
  \bibfield  {author} {\bibinfo {author} {\bibfnamefont {V.~M.}\ \bibnamefont
  {Braun}}, \bibinfo {author} {\bibfnamefont {M.}~\bibnamefont {G\"ockeler}},
  \bibinfo {author} {\bibfnamefont {R.}~\bibnamefont {Horsley}}, \bibinfo
  {author} {\bibfnamefont {T.}~\bibnamefont {Kaltenbrunner}}, \bibinfo {author}
  {\bibfnamefont {A.}~\bibnamefont {Lenz}}, \bibinfo {author} {\bibfnamefont
  {Y.}~\bibnamefont {Nakamura}}, \bibinfo {author} {\bibfnamefont
  {D.}~\bibnamefont {Pleiter}}, \bibinfo {author} {\bibfnamefont {P.~E.~L.}\
  \bibnamefont {Rakow}}, \bibinfo {author} {\bibfnamefont {J.}~\bibnamefont
  {Rohrwild}}, \bibinfo {author} {\bibfnamefont {A.}~\bibnamefont {Sch\"afer}},
  \bibinfo {author} {\bibfnamefont {G.}~\bibnamefont {Schierholz}}, \bibinfo
  {author} {\bibfnamefont {H.}~\bibnamefont {St\"uben}}, \bibinfo {author}
  {\bibfnamefont {N.}~\bibnamefont {Warkentin}},\ and\ \bibinfo {author}
  {\bibfnamefont {J.~M.}\ \bibnamefont {Zanotti}},\ }\href
  {https://doi.org/10.1103/PhysRevLett.103.072001} {\bibfield  {journal}
  {\bibinfo  {journal} {Phys. Rev. Lett.}\ }\textbf {\bibinfo {volume} {103}},\
  \bibinfo {pages} {072001} (\bibinfo {year} {2009})}\BibitemShut {NoStop}%
\bibitem [{\citenamefont {Anikin}\ \emph {et~al.}(2015)\citenamefont {Anikin},
  \citenamefont {Braun},\ and\ \citenamefont {Offen}}]{PhysRevD.92.014018}%
  \BibitemOpen
  \bibfield  {author} {\bibinfo {author} {\bibfnamefont {I.~V.}\ \bibnamefont
  {Anikin}}, \bibinfo {author} {\bibfnamefont {V.~M.}\ \bibnamefont {Braun}},\
  and\ \bibinfo {author} {\bibfnamefont {N.}~\bibnamefont {Offen}},\ }\href
  {https://doi.org/10.1103/PhysRevD.92.014018} {\bibfield  {journal} {\bibinfo
  {journal} {Phys. Rev. D}\ }\textbf {\bibinfo {volume} {92}},\ \bibinfo
  {pages} {014018} (\bibinfo {year} {2015})}\BibitemShut {NoStop}%
\bibitem [{\citenamefont {Gutsche}\ \emph {et~al.}(2020)\citenamefont
  {Gutsche}, \citenamefont {Lyubovitskij},\ and\ \citenamefont
  {Schmidt}}]{PhysRevD.101.034026}%
  \BibitemOpen
  \bibfield  {author} {\bibinfo {author} {\bibfnamefont {T.}~\bibnamefont
  {Gutsche}}, \bibinfo {author} {\bibfnamefont {V.~E.}\ \bibnamefont
  {Lyubovitskij}},\ and\ \bibinfo {author} {\bibfnamefont {I.}~\bibnamefont
  {Schmidt}},\ }\href {https://doi.org/10.1103/PhysRevD.101.034026} {\bibfield
  {journal} {\bibinfo  {journal} {Phys. Rev. D}\ }\textbf {\bibinfo {volume}
  {101}},\ \bibinfo {pages} {034026} (\bibinfo {year} {2020})}\BibitemShut
  {NoStop}%
\bibitem [{\citenamefont {Lyubovitskij}\ and\ \citenamefont
  {Schmidt}(2020)}]{PhysRevD.102.094008}%
  \BibitemOpen
  \bibfield  {author} {\bibinfo {author} {\bibfnamefont {V.~E.}\ \bibnamefont
  {Lyubovitskij}}\ and\ \bibinfo {author} {\bibfnamefont {I.}~\bibnamefont
  {Schmidt}},\ }\href {https://doi.org/10.1103/PhysRevD.102.094008} {\bibfield
  {journal} {\bibinfo  {journal} {Phys. Rev. D}\ }\textbf {\bibinfo {volume}
  {102}},\ \bibinfo {pages} {094008} (\bibinfo {year} {2020})}\BibitemShut
  {NoStop}%
\bibitem [{\citenamefont {Burkert}\ and\ \citenamefont
  {Lee}(2004)}]{Burkert:2004sk}%
  \BibitemOpen
  \bibfield  {author} {\bibinfo {author} {\bibfnamefont {V.~D.}\ \bibnamefont
  {Burkert}}\ and\ \bibinfo {author} {\bibfnamefont {T.~S.~H.}\ \bibnamefont
  {Lee}},\ }\href {https://doi.org/10.1142/S0218301304002545} {\bibfield
  {journal} {\bibinfo  {journal} {Int. J. Mod. Phys. E}\ }\textbf {\bibinfo
  {volume} {13}},\ \bibinfo {pages} {1035} (\bibinfo {year} {2004})},\ \Eprint
  {https://arxiv.org/abs/nucl-ex/0407020} {arXiv:nucl-ex/0407020} \BibitemShut
  {NoStop}%
\bibitem [{\citenamefont {Kamano}\ \emph {et~al.}(2013)\citenamefont {Kamano},
  \citenamefont {Nakamura}, \citenamefont {Lee},\ and\ \citenamefont
  {Sato}}]{PhysRevC.88.035209}%
  \BibitemOpen
  \bibfield  {author} {\bibinfo {author} {\bibfnamefont {H.}~\bibnamefont
  {Kamano}}, \bibinfo {author} {\bibfnamefont {S.~X.}\ \bibnamefont
  {Nakamura}}, \bibinfo {author} {\bibfnamefont {T.-S.~H.}\ \bibnamefont
  {Lee}},\ and\ \bibinfo {author} {\bibfnamefont {T.}~\bibnamefont {Sato}},\
  }\href {https://doi.org/10.1103/PhysRevC.88.035209} {\bibfield  {journal}
  {\bibinfo  {journal} {Phys. Rev. C}\ }\textbf {\bibinfo {volume} {88}},\
  \bibinfo {pages} {035209} (\bibinfo {year} {2013})}\BibitemShut {NoStop}%
\bibitem [{\citenamefont {Jido}\ \emph {et~al.}(2008)\citenamefont {Jido},
  \citenamefont {D\"oring},\ and\ \citenamefont {Oset}}]{PhysRevC.77.065207}%
  \BibitemOpen
  \bibfield  {author} {\bibinfo {author} {\bibfnamefont {D.}~\bibnamefont
  {Jido}}, \bibinfo {author} {\bibfnamefont {M.}~\bibnamefont {D\"oring}},\
  and\ \bibinfo {author} {\bibfnamefont {E.}~\bibnamefont {Oset}},\ }\href
  {https://doi.org/10.1103/PhysRevC.77.065207} {\bibfield  {journal} {\bibinfo
  {journal} {Phys. Rev. C}\ }\textbf {\bibinfo {volume} {77}},\ \bibinfo
  {pages} {065207} (\bibinfo {year} {2008})}\BibitemShut {NoStop}%
\bibitem [{\citenamefont {Ramalho}\ and\ \citenamefont
  {Pe\~na}(2014)}]{PhysRevD.89.094016}%
  \BibitemOpen
  \bibfield  {author} {\bibinfo {author} {\bibfnamefont {G.}~\bibnamefont
  {Ramalho}}\ and\ \bibinfo {author} {\bibfnamefont {M.~T.}\ \bibnamefont
  {Pe\~na}},\ }\href {https://doi.org/10.1103/PhysRevD.89.094016} {\bibfield
  {journal} {\bibinfo  {journal} {Phys. Rev. D}\ }\textbf {\bibinfo {volume}
  {89}},\ \bibinfo {pages} {094016} (\bibinfo {year} {2014})}\BibitemShut
  {NoStop}%
\bibitem [{\citenamefont {Ramalho}\ and\ \citenamefont
  {Pe\~na}(2017)}]{PhysRevD.95.014003}%
  \BibitemOpen
  \bibfield  {author} {\bibinfo {author} {\bibfnamefont {G.}~\bibnamefont
  {Ramalho}}\ and\ \bibinfo {author} {\bibfnamefont {M.~T.}\ \bibnamefont
  {Pe\~na}},\ }\href {https://doi.org/10.1103/PhysRevD.95.014003} {\bibfield
  {journal} {\bibinfo  {journal} {Phys. Rev. D}\ }\textbf {\bibinfo {volume}
  {95}},\ \bibinfo {pages} {014003} (\bibinfo {year} {2017})}\BibitemShut
  {NoStop}%
\bibitem [{\citenamefont {Ramalho}\ \emph {et~al.}(2012)\citenamefont
  {Ramalho}, \citenamefont {Jido},\ and\ \citenamefont
  {Tsushima}}]{PhysRevD.85.093014}%
  \BibitemOpen
  \bibfield  {author} {\bibinfo {author} {\bibfnamefont {G.}~\bibnamefont
  {Ramalho}}, \bibinfo {author} {\bibfnamefont {D.}~\bibnamefont {Jido}},\ and\
  \bibinfo {author} {\bibfnamefont {K.}~\bibnamefont {Tsushima}},\ }\href
  {https://doi.org/10.1103/PhysRevD.85.093014} {\bibfield  {journal} {\bibinfo
  {journal} {Phys. Rev. D}\ }\textbf {\bibinfo {volume} {85}},\ \bibinfo
  {pages} {093014} (\bibinfo {year} {2012})}\BibitemShut {NoStop}%
\bibitem [{\citenamefont {Julia-Diaz}\ \emph {et~al.}(2008)\citenamefont
  {Julia-Diaz}, \citenamefont {Lee}, \citenamefont {Matsuyama}, \citenamefont
  {Sato},\ and\ \citenamefont {Smith}}]{JuliaDiaz:2007fa}%
  \BibitemOpen
  \bibfield  {author} {\bibinfo {author} {\bibfnamefont {B.}~\bibnamefont
  {Julia-Diaz}}, \bibinfo {author} {\bibfnamefont {T.~S.~H.}\ \bibnamefont
  {Lee}}, \bibinfo {author} {\bibfnamefont {A.}~\bibnamefont {Matsuyama}},
  \bibinfo {author} {\bibfnamefont {T.}~\bibnamefont {Sato}},\ and\ \bibinfo
  {author} {\bibfnamefont {L.~C.}\ \bibnamefont {Smith}},\ }\href
  {https://doi.org/10.1103/PhysRevC.77.045205} {\bibfield  {journal} {\bibinfo
  {journal} {Phys. Rev. C}\ }\textbf {\bibinfo {volume} {77}},\ \bibinfo
  {pages} {045205} (\bibinfo {year} {2008})},\ \Eprint
  {https://arxiv.org/abs/0712.2283} {arXiv:0712.2283 [nucl-th]} \BibitemShut
  {NoStop}%
\bibitem [{\citenamefont {An}\ and\ \citenamefont {Zou}(2009)}]{An2009}%
  \BibitemOpen
  \bibfield  {author} {\bibinfo {author} {\bibfnamefont {C.~S.}\ \bibnamefont
  {An}}\ and\ \bibinfo {author} {\bibfnamefont {B.~S.}\ \bibnamefont {Zou}},\
  }\href {https://doi.org/10.1140/epja/i2008-10698-x} {\bibfield  {journal}
  {\bibinfo  {journal} {The European Physical Journal A}\ }\textbf {\bibinfo
  {volume} {39}},\ \bibinfo {pages} {195} (\bibinfo {year} {2009})}\BibitemShut
  {NoStop}%
\bibitem [{\citenamefont {Zou}(2010)}]{ZOU2010199}%
  \BibitemOpen
  \bibfield  {author} {\bibinfo {author} {\bibfnamefont {B.-S.}\ \bibnamefont
  {Zou}},\ }\href
  {https://doi.org/https://doi.org/10.1016/j.nuclphysa.2010.01.194} {\bibfield
  {journal} {\bibinfo  {journal} {Nuclear Physics A}\ }\textbf {\bibinfo
  {volume} {835}},\ \bibinfo {pages} {199} (\bibinfo {year}
  {2010})}\BibitemShut {NoStop}%
\bibitem [{\citenamefont {Kaewsnod}\ \emph
  {et~al.}(2022{\natexlab{a}})\citenamefont {Kaewsnod}, \citenamefont {Xu},
  \citenamefont {Zhao}, \citenamefont {Liu}, \citenamefont {Srisuphaphon},
  \citenamefont {Limphirat},\ and\ \citenamefont {Yan}}]{PhysRevD.105.016008}%
  \BibitemOpen
  \bibfield  {author} {\bibinfo {author} {\bibfnamefont {A.}~\bibnamefont
  {Kaewsnod}}, \bibinfo {author} {\bibfnamefont {K.}~\bibnamefont {Xu}},
  \bibinfo {author} {\bibfnamefont {Z.}~\bibnamefont {Zhao}}, \bibinfo {author}
  {\bibfnamefont {X.~Y.}\ \bibnamefont {Liu}}, \bibinfo {author} {\bibfnamefont
  {S.}~\bibnamefont {Srisuphaphon}}, \bibinfo {author} {\bibfnamefont
  {A.}~\bibnamefont {Limphirat}},\ and\ \bibinfo {author} {\bibfnamefont
  {Y.}~\bibnamefont {Yan}},\ }\href
  {https://doi.org/10.1103/PhysRevD.105.016008} {\bibfield  {journal} {\bibinfo
   {journal} {Phys. Rev. D}\ }\textbf {\bibinfo {volume} {105}},\ \bibinfo
  {pages} {016008} (\bibinfo {year} {2022}{\natexlab{a}})}\BibitemShut
  {NoStop}%
\bibitem [{\citenamefont {Kaewsnod}\ \emph
  {et~al.}(2022{\natexlab{b}})\citenamefont {Kaewsnod}, \citenamefont {Xu},
  \citenamefont {Zhao}, \citenamefont {Liu}, \citenamefont {Srisuphaphon},
  \citenamefont {Limphirat},\ and\ \citenamefont {Yan}}]{MosEpja}%
  \BibitemOpen
  \bibfield  {author} {\bibinfo {author} {\bibfnamefont {A.}~\bibnamefont
  {Kaewsnod}}, \bibinfo {author} {\bibfnamefont {K.}~\bibnamefont {Xu}},
  \bibinfo {author} {\bibfnamefont {Z.}~\bibnamefont {Zhao}}, \bibinfo {author}
  {\bibfnamefont {X.}~\bibnamefont {Liu}}, \bibinfo {author} {\bibfnamefont
  {S.}~\bibnamefont {Srisuphaphon}}, \bibinfo {author} {\bibfnamefont
  {A.}~\bibnamefont {Limphirat}},\ and\ \bibinfo {author} {\bibfnamefont
  {Y.}~\bibnamefont {Yan}},\ }\href
  {https://doi.org/10.1140/epja/s10050-022-00837-0} {\bibfield  {journal}
  {\bibinfo  {journal} {The European Physical Journal A}\ }\textbf {\bibinfo
  {volume} {58}},\ \bibinfo {pages} {185} (\bibinfo {year}
  {2022}{\natexlab{b}})}\BibitemShut {NoStop}%
\bibitem [{\citenamefont {Xu}\ \emph {et~al.}(2020)\citenamefont {Xu},
  \citenamefont {Kaewsnod}, \citenamefont {Zhao}, \citenamefont {Liu},
  \citenamefont {Srisuphaphon}, \citenamefont {Limphirat},\ and\ \citenamefont
  {Yan}}]{Kai2020PRD}%
  \BibitemOpen
  \bibfield  {author} {\bibinfo {author} {\bibfnamefont {K.}~\bibnamefont
  {Xu}}, \bibinfo {author} {\bibfnamefont {A.}~\bibnamefont {Kaewsnod}},
  \bibinfo {author} {\bibfnamefont {Z.}~\bibnamefont {Zhao}}, \bibinfo {author}
  {\bibfnamefont {X.~Y.}\ \bibnamefont {Liu}}, \bibinfo {author} {\bibfnamefont
  {S.}~\bibnamefont {Srisuphaphon}}, \bibinfo {author} {\bibfnamefont
  {A.}~\bibnamefont {Limphirat}},\ and\ \bibinfo {author} {\bibfnamefont
  {Y.}~\bibnamefont {Yan}},\ }\href
  {https://doi.org/10.1103/PhysRevD.101.076025} {\bibfield  {journal} {\bibinfo
   {journal} {Phys. Rev. D}\ }\textbf {\bibinfo {volume} {101}},\ \bibinfo
  {pages} {076025} (\bibinfo {year} {2020})}\BibitemShut {NoStop}%
\bibitem [{\citenamefont {Xu}\ \emph {et~al.}(2014)\citenamefont {Xu},
  \citenamefont {Ritjoho}, \citenamefont {Srisuphaphon},\ and\ \citenamefont
  {Yan}}]{Kai2014}%
  \BibitemOpen
  \bibfield  {author} {\bibinfo {author} {\bibfnamefont {K.}~\bibnamefont
  {Xu}}, \bibinfo {author} {\bibfnamefont {N.}~\bibnamefont {Ritjoho}},
  \bibinfo {author} {\bibfnamefont {S.}~\bibnamefont {Srisuphaphon}},\ and\
  \bibinfo {author} {\bibfnamefont {Y.}~\bibnamefont {Yan}},\ }\href
  {https://doi.org/10.1142/S2010194514602518} {\bibfield  {journal} {\bibinfo
  {journal} {International Journal of Modern Physics: Conference Series}\
  }\textbf {\bibinfo {volume} {29}},\ \bibinfo {pages} {1460251} (\bibinfo
  {year} {2014})},\ \Eprint
  {https://arxiv.org/abs/https://doi.org/10.1142/S2010194514602518}
  {https://doi.org/10.1142/S2010194514602518} \BibitemShut {NoStop}%
\bibitem [{\citenamefont {Xu}\ \emph {et~al.}(2019)\citenamefont {Xu},
  \citenamefont {Kaewsnod}, \citenamefont {Liu}, \citenamefont {Srisuphaphon},
  \citenamefont {Limphirat},\ and\ \citenamefont {Yan}}]{PhysRevC.100.065207}%
  \BibitemOpen
  \bibfield  {author} {\bibinfo {author} {\bibfnamefont {K.}~\bibnamefont
  {Xu}}, \bibinfo {author} {\bibfnamefont {A.}~\bibnamefont {Kaewsnod}},
  \bibinfo {author} {\bibfnamefont {X.~Y.}\ \bibnamefont {Liu}}, \bibinfo
  {author} {\bibfnamefont {S.}~\bibnamefont {Srisuphaphon}}, \bibinfo {author}
  {\bibfnamefont {A.}~\bibnamefont {Limphirat}},\ and\ \bibinfo {author}
  {\bibfnamefont {Y.}~\bibnamefont {Yan}},\ }\href
  {https://doi.org/10.1103/PhysRevC.100.065207} {\bibfield  {journal} {\bibinfo
   {journal} {Phys. Rev. C}\ }\textbf {\bibinfo {volume} {100}},\ \bibinfo
  {pages} {065207} (\bibinfo {year} {2019})}\BibitemShut {NoStop}%
\bibitem [{\citenamefont {Navas}\ \emph {et~al.}(2024)\citenamefont {Navas}
  \emph {et~al.}}]{PDG2024}%
  \BibitemOpen
  \bibfield  {author} {\bibinfo {author} {\bibfnamefont {S.}~\bibnamefont
  {Navas}} \emph {et~al.} (\bibinfo {collaboration} {Particle Data Group}),\
  }\href {https://doi.org/10.1103/PhysRevD.110.030001} {\bibfield  {journal}
  {\bibinfo  {journal} {Phys. Rev. D}\ }\textbf {\bibinfo {volume} {110}},\
  \bibinfo {pages} {030001} (\bibinfo {year} {2024})}\BibitemShut {NoStop}%
\end{thebibliography}%

\end{document}